\documentclass[twocolumn]{aastex63}

\shorttitle{Massive Young Stellar Objects in the Galactic Center.\ II.}
\shortauthors{Jang et al.}

\begin{document}

\title{Massive Young Stellar Objects in the Galactic Center.\ II.\ Seeing Through the Ice-rich Envelopes}

\author{Dajeong Jang}
\altaffiliation{Current Address: Department of Physics and Astronomy, Seoul National University, Seoul, Republic of Korea.}
\affiliation{Department of Science Education, Ewha Womans University, 52 Ewhayeodae-gil, Seodaemun-gu, Seoul 03760, Republic of Korea; deokkeun@ewha.ac.kr}

\author{Deokkeun An}
\altaffiliation{Visiting Astronomer at the Infrared Telescope Facility, which is operated by the University of Hawaii under contract 80HQTR19D0030 with the National Aeronautics and Space Administration.}
\affiliation{Department of Science Education, Ewha Womans University, 52 Ewhayeodae-gil, Seodaemun-gu, Seoul 03760, Korea; deokkeun@ewha.ac.kr}

\author{Kris Sellgren}
\altaffiliation{Visiting Astronomer at the Infrared Telescope Facility, which is operated by the University of Hawaii under contract 80HQTR19D0030 with the National Aeronautics and Space Administration.}
\affiliation{Department of Astronomy, Ohio State University, 140 West 18th Avenue, Columbus, OH 43210, USA}

\author{Solange V. Ram{\'{\i}}rez}
\altaffiliation{Visiting Astronomer at the Infrared Telescope Facility, which is operated by the University of Hawaii under contract 80HQTR19D0030 with the National Aeronautics and Space Administration.}
\affiliation{Carnegie Observatories, 813 Santa Barbara Street, Pasadena, CA 91101, USA }

\author{A. C. Adwin Boogert}
\affiliation{Institute for Astronomy, University of Hawaii, 2680 Woodlawn Dr., Honolulu, HI 98622, USA}

\author{Mathias Schultheis}
\affiliation{Universit{\'{e}} C{\^{o}}te d'Azur, Observatoire de la C{\^{o}}te d’Azur, CNRS, Laboratoire Lagrange, Blvd de l'Observatoire, 06304 Nice, France}

\accepted{for publication in the Astrophysical Journal}

\begin{abstract}

To study the demographics of interstellar ices in the Central Molecular Zone (CMZ) of the Milky Way, we obtain near-infrared spectra of $109$ red point sources using NASA IRTF/SpeX at Maunakea. We select the sample from near- and mid-infrared photometry, including $12$ objects in the previous paper of this series, to ensure that these sources trace a large amount of absorption through clouds in each line of sight. We find that most of the sample ($100$ objects) show CO band-head absorption at $2.3\ \micron$, tagging them as red (super-) giants. Despite the photospheric signature, however, a fraction of the sample with $L$-band spectra ($9/82=0.11$) exhibit large H$_2$O ice column densities ($N > 2\times10^{18}\ {\rm cm}^{-2}$), and six of them also reveal CH$_3$OH ice absorption. As one of such objects is identified as a young stellar object (YSO) in our previous work, these ice-rich sight lines are likely associated with background stars in projection to an extended envelope of a YSO or a dense cloud core. The low frequency of such objects in the early stage of stellar evolution implies a low star-formation rate ($\la0.02\ M_\odot$ yr$^{-1}$), reinforcing the previous claim on the suppressed star-formation activity in the CMZ. Our data also indicate that the strong ``shoulder'' CO$_2$ ice absorption at $15.4\ \micron$ observed in YSO candidates in the previous paper arises from CH$_3$OH-rich ice grains having a large CO$_2$ concentration [$N {\rm (CO_2)} / N {\rm (CH_3OH)} \approx 1/3$].

\end{abstract}


\section{Introduction}\label{sec:intro}

The Central Molecular Zone \citep[CMZ; see][and references therein]{morris:96}\footnote{Throughout the paper, we refer the Galactic center (GC) to a central parsec region around Sgr~A$^*$, unless otherwise specified.} is a massive disk (and torus) of molecular gas in the Milky Way's center, containing approximately $10\%$ of the molecular clouds in the Galaxy within a Galactocentric radius of just $\sim250$~pc. At spatial scales unapproachable for other galaxies, it is a unique window to study stellar systems and star-forming activity in galactic nuclei. As its large molecular mass \citep[$\sim1$--$6\times10^7\ M_\odot$;][]{ferriere:07} suggests, the system exhibits several signposts of star formation, including compact and ultra-compact \ion{H}{2} regions \citep[e.g.,][]{gaume:95}, massive stars in the field and star clusters \citep[e.g.,][]{mauerhan:10,dong:12}, H$_2$O/CH$_3$OH masers \citep[e.g.,][]{caswell:10,longmore:17}, and high-density clumps \citep[e.g.,][]{rathborne:14,kauffmann:17}. However, recent studies indicate that star-forming activities in the CMZ are at least an order of magnitude lower than expected despite the large mass of its dense gas clouds \citep[e.g.,][]{longmore:13,kruijssen:14,kauffmann:17}. An inefficient conversion of gas into stars can be related to the unique and extreme physical conditions of the CMZ \citep[see][and references therein]{mills:17}; in this region, clouds have high average gas temperatures ($50$--$100$~K), strong magnetic fields ($\sim1$~milli-gauss), and supersonic turbulence (FWHM$\sim20$--$50\ {\rm km\ s^{-1}}$).

Among several tracers of star formation, young stellar objects (YSOs) serve as a useful indicator of on-going star formation, complementing observational constraints provided by other tracers. Based on observations using the Infrared Spectrograph 
\citep[IRS;][]{houck:04} onboard the Spitzer Space Telescope \citep{werner:04}, \citet[][hereafter Paper~I, see also \citealt{an:09}]{an:11} identified $35$ YSOs in the CMZ based on the presence of molecular gas and ice absorptions. These IRS spectra reveal gas-phase absorption from C$_2$H$_2$ ($13.7\ \micron$), HCN ($14.0\ \micron$), and CO$_2$ ($15.0\ \micron$), which trace warm and dense gas in the circumstellar disk and/or YSO envelopes \citep[e.g.,][]{lahuis:00,boonman:03}. In addition, they exhibit a strong and wide absorption profile of CO$_2$ ice at $15\ \micron$. The absorption at its short-wavelength side can be attributed to a combination of CO$_2$ in H$_2$O-rich (polar; $\sim15.3\ \micron$) and CO-rich (apolar; $\sim15.1\ \micron$) ice mantles, as frequently seen toward dark cloud cores and YSOs, including lines of sight toward Sgr~A$^*$ \citep{gerakines:99}. On the other hand, many of the IRS spectra show excess absorption in the long-wavelength wing centered at $15.4\ \micron$, which was postulated to form by Lewis acid-base interactions of CO$_2$ with CH$_3$OH \citep{dartois:99a}. Indeed, a large amount of solid CH$_3$OH was also found in one of the Spitzer/IRS YSO  \citep{an:17}, as seen toward other massive YSOs in the Milky Way's disk \citep{dartois:99b}.

The mid-infrared (IR) spectroscopic observations in Paper~I demonstrate the youth of these objects, and thereby provide a unique sample of a YSO population in the CMZ. They can also serve as a window to probe ice compositions in and around a massive YSO in the CMZ and study complex chemical reactions at work between gas and ice grains. In particular, CO$_2$ ice absorption at $15\ \micron$ provides the most tantalizing aspect of the physical and chemical properties of the CMZ clouds. Other than this sample, a strong and wide $15\ \micron$ CO$_2$ ice profile has only been observed toward about half a dozen massive YSOs in the Galactic disk \citep{gerakines:99,boogert:15}, of which the strength of the shoulder absorption is comparable to those from other CO$_2$ components. The observed CO$_2$ profiles of YSOs and possible YSOs in Paper~I look quite similar to those seen toward these objects. The shoulder CO$_2$ absorptions are also observed toward low-mass YSOs in the disk, but their relative strengths are significantly weaker \citep{pontoppidan:08}.

Although the above result may simply reflect active star formation in the CMZ, a factor of about five increase in the number of such objects raises the question of whether the formation of CH$_3$OH-rich ice grains are somehow enhanced in this region by its unique physical and chemical conditions. In an effort to answer such questions, here we undertook $2$--$4\ \micron$ spectroscopic observations for a large number of CMZ sources, with a goal of providing extensive knowledge on the demographics of ices in this region. A wealth of molecular ice absorption features have been observed toward bright, compact sources in the central parsec of the Galaxy \citep{mcfadzean:89,lutz:96,chiar:00,gibb:04,moultaka:15}, yet no systematic studies on ices have been conducted in a much larger volume across the CMZ. The observed spectra cover a broad absorption band from the O--H stretching mode of H$_2$O ice, so they complement Spitzer/IRS spectra in Paper~I, from which only a weak constraint could be derived from its libration band.

Nonetheless, it is not trivial to interpret spectra of CMZ sources, since any observed spectroscopic features represent an accumulation of absorptions from clouds that are located along these complex lines of sight \citep[e.g.,][]{martin:04,geballe:21}. The accumulation of dust from these cloud complexes also produces an excessive and variable amount of visual extinction of the order of $\langle A_{\rm V} \rangle\sim30$~mag \citep[e.g.,][]{schultheis:09,nogueraslara:21}. Earlier observations \citep[e.g.,][]{lebofsky:79} revealed that molecular clouds are responsible for at least parts of the dust extinction toward the GC. In particular, \citet{mcfadzean:89} demonstrated, based on $L$-band spectroscopy, that the foreground absorption has two components: a uniform absorption from diffuse interstellar medium (ISM) as traced by the $3.4\ \micron$ band from hydrocarbons and a spatially variable absorption from intervening molecular clouds as traced by $3\ \micron$ H$_2$O ice absorption. \citet{whittet:97} further made an order-of-magnitude estimate of the contribution from each of the dust components based on gas-phase CO and $3\ \micron$ H$_2$O ice observations, and concluded that about one-third of the total extinction ($A_{\rm V} \approx 10$ mag) is associated with refractory dusts in intervening molecular clouds, while the rest ($A_{\rm V} \approx 20$ mag) originates from the foreground diffuse ISM.

In addition to foreground disk clouds and large-scale cloud complexes in the CMZ, there can be a local enhancement of gas, such as a dense cloud core or an extended envelope of a massive YSO. Because the extinction is strongly variable across the region owing to patchy foreground clouds, high extinction is not a sufficient condition to guarantee the youth of objects, and red IR colors are not always enough to distinguish such young objects from the vast majority of old stars behind extra dust clouds. Under such circumstances, pure methanol ices can be used as a powerful tracer for isolating the highest density regions in the CMZ, where massive star formation is imminent or likely underway. Methanol ices form efficiently in dense ($\ga10^5\ {\rm cm}^{-3}$) and cold ($\la15$~K) environments \citep{watanabe:03,cuppen:09,coutens:17} and serve as a key molecular species in the formation of more complex organic molecules observed in the CMZ \citep[e.g.,][]{requena-torres:06,requena-torres:08}. In the Galactic disk, CH$_3$OH ices are found in various environments, including YSOs \citep{allamandola:92,skinner:92,brooke:99,dartois:99b,pontoppidan:03,pontoppidan:04,whittet:11}, isolated dense cores \citep{boogert:11,boogert:15} and quiescent clouds \citep{chiar:96,chiar:11}. Within each category, the abundances of solid CH$_3$OH relative to water ice vary significantly from less than $1\%$ to $12\%$ in dense clouds and cloud cores, and up to $30\%$ in YSOs \citep{boogert:15}. In contrast, CH$_3$OH ices have never been detected in diffuse ISM. In previous studies, upper limits have been measured toward Sgr~A$^*$ ($< 4\%$) and the Quintuplet cluster star ($<27\%$) \citep{chiar:00,gibb:04,moultaka:15}. In this regard, the detection of the C--H stretching mode of CH$_3$OH at $3.54\ \micron$ toward one of the Spitzer/IRS sources in Paper~I (SSTGC~726327) and its large abundance relative to solid H$_2$O ($17\pm3\%$) provide a tantalizing evidence that methanol-rich ice grains are localized in the highest density regions in the CMZ \citep{an:17}.

In this paper, we present $2$--$4\ \micron$ spectra of $109$ objects in the CMZ, some of which are taken from Paper~I. The $3\ \micron$ H$_2$O ice and $3.54\ \micron$ CH$_3$OH bands are primarily used to probe ice-rich lines of sight toward the CMZ. Their column densities are also combined with information on individual CO$_2$ ice components in Paper~I to establish a link between the CO$_2$ shoulder absorption and CH$_3$OH on a firmer ground. This paper is organized as follows. We describe our sample selections and observations in \S~\ref{sec:obs}. In \S~\ref{sec:analysis}, we provide detailed descriptions on how we extracted information on molecular and gaseous absorptions from the spectra. In \S~\ref{sec:result}, we present the observed distributions of ices and dusts in the CMZ as traced by our sample of red point-like objects. We summarize our results in \S~\ref{sec:summary}.

\section{Observations and Data Reduction}\label{sec:obs}

We obtained $2$--$4\ \micron$ spectra for $109$ red point-like sources using SpeX, a medium-resolution ($R\sim2000$--$2500$) cross-dispersed spectrograph and imager \citep{rayner:03} at the 3.0-m NASA Infrared Telescope Facility (IRTF) on Maunakea. For these objects that are potentially associated with the CMZ (as judged by red IR colors from heavy foreground dust extinction; see below), we carried observing runs on UT dates 2009 May 14--15 and 17, 2011 July 17--20, 2017 May 28--30, June 21--23, and July 23--24 to collect spectra for $11$, $28$, and $70$ science targets in 2009, 2011, and 2017, respectively. In 2009 and 2017, we mainly used the Long Cross-dispersed Mode ($1.67$--$4.2\ \micron$) with a $15\arcsec$ long slit to obtain both $K$- and $L$-band spectra. For the observing targets in 2011, we used the Short Cross-dispersed Mode (SXD, $0.70$--$2.55\ \micron$) to obtain their $K$-band spectra only. There has been a major system upgrade in SpeX in 2014 that improved the wavelength coverage and sensitivity, including an installation of a larger format CCD ($2048\times2048$ array as compared to the earlier $1024\times1024$ array). Otherwise, we kept the same instrument set-up throughout the entire observing runs. We used the $0.5\arcsec$--$0.8\arcsec$ slit to match the seeing condition. Most data were taken at air mass less than $2$ under clear or fair sky conditions with varying amounts of thin cirrus.

\subsection{Observing Run in 2009}

\begin{deluxetable}{lccccc}
\tablecaption{Properties of the Sample from Paper~I\label{tab:a11}}
\tabletypesize{\scriptsize}
\tablehead{
   \colhead{SSTGC} &
   \colhead{YSO\tablenotemark{\scriptsize a}} & 
   \colhead{CO abs} &
   \colhead{Solid H$_2$O\tablenotemark{\scriptsize b}} &  
   \colhead{Solid CH$_3$OH} & 
   \colhead{note}
}
\startdata
300758  & maybe & maybe & weak     &     & low SNR \\
348392  &    &  & moderate &     & \\
388790  &    &  & weak     &     & \\
404312  &    & maybe & weak     &     & \\
405235  &    &  & moderate &     & \\
610642  & maybe &  & strong   &     & low SNR \\
653270  & maybe & \checkmark & strong   & \checkmark   & \\
696367  &    & \checkmark & strong   & \checkmark   & \\
716531  &    &  & weak     &     & \\
726327E &  YSO  & \checkmark & strong   & \checkmark   & \\
726327W & YSO &  & strong   & maybe & low SNR \\
817031  &    & \checkmark & moderate &     & \\
\enddata
\tablenotetext{a}{YSO classification in Paper~I.}
\tablenotetext{b}{Strong: $\tau_{3.0} > 1.2$ [$N$(H$_2$O) $> 2 \times 10^{18}$ cm$^{-2}$]. Moderate: $0.5 < \tau_{3.0} \leq 1.2$.}
\end{deluxetable}

\clearpage
\startlongtable
\begin{deluxetable*}{ccccccc}
\tablecaption{IRTF/SpeX Targets\label{tab:sample}}
\tabletypesize{\scriptsize}             
\tablehead{
\colhead{SSTGC} &
\colhead{2MASS Name}&
\colhead{$l$} &
\colhead{$b$} &
\colhead{Date of obs.} &
\colhead{Bands}  &
\colhead{Note} \\
\colhead{} &
\colhead{}&
\colhead{(deg)} &
\colhead{(deg)} &
\colhead{} &
\colhead{}  &
\colhead{}
}
\startdata
131902	&	17424354-2949094	&	358.9174	&	0.0748	&	 2011 Jul 19 	&	K	&		\\
153970	&	17425720-2955157	&	358.8569	&	-0.0207	&	 2011 Jul 19 	&	K	&		\\
158600	&	17430001-2938476	&	359.0957	&	0.1150	&	 2017 Jun 21 	&	K,L	&		\\
183877	&	17431543-2924587	&	359.3207	&	0.1884	&	 2017 May 29 	&	K,L	&		\\
188670	&	17431822-2935397	&	359.1748	&	0.0863	&	 2017 Jul 23 	&	K,L	&		\\
203189	&	17432603-2947331	&	359.0210	&	-0.0419	&	 2017 May 29 	&	K,L	&		\\
207109	&	17432817-2917418	&	359.4483	&	0.2128	&	 2017 May 29 	&	K,L	&		\\
219495	&	17433479-2940304	&	359.1376	&	-0.0072	&	 2017 May 28 	&	K,L	&		\\
220156	&	17433512-2924472	&	359.3612	&	0.1293	&	 2017 May 28 	&	K,L	&		\\
226380	&	17433844-2952527	&	358.9689	&	-0.1266	&	 2017 Jun 22 	&	K,L	&		\\
239331	&	17434525-2953193	&	358.9755	&	-0.1515	&	 2017 Jun 21 	&	K,L	&		\\
258517	&	17435477-2946031	&	359.0968	&	-0.1173	&	 2011 Jul 19 	&	K	&		\\
261714	&	17435631-2936031	&	359.2416	&	-0.0347	&	 2017 May 30 	&	K,L	&		\\
263790	&	17435729-2931465	&	359.3042	&	-0.0003	&	 2017 May 30 	&	K,L	&		\\
265318	&	17435803-2943121	&	359.1434	&	-0.1024	&	 2017 May 28 	&	K,L	&		\\
268003	&	17435926-2919517	&	359.4771	&	0.0976	&	 2017 May 28 	&	K,L	&		\\
272863	&	\nodata	&	359.6135	&	0.1719	&	2017 May 28	&	K,L	&		\\
275827	&	17440298-2838455	&	359.2791	&	-0.0397	&	 2011 Jul 19 	&	K	&		\\
278131	&	17440402-2926225	&	359.3937	&	0.0260	&	 2017 Jun 22 	&	K,L	&		\\
281221	&	17440547-2912375	&	359.5916	&	0.1416	&	 2017 Jun 22 	&	K,L	&		\\
285458	&	17440744-2927376	&	359.3824	&	0.0046	&	 2017 May 29 	&	K,L	&		\\
286731	&	17440804-2932225	&	359.3161	&	-0.0387	&	 2017 May 29 	&	K,L	&		\\
296989	&	17441277-2926555	&	359.4025	&	-0.0058	&	 2017 May 29 	&	K,L	&		\\
298405	&	17441342-2915408	&	359.5634	&	0.0903	&	 2017 Jun 21 	&	K,L	&		\\
300758	&	17441448-2923220	&	359.4562	&	0.0199	&	 2009 May 14 	&	K,L	&		\\
306306	&	17441690-2938193	&	359.2485	&	-0.1180	&	 2017 May 30 	&	K,L	&		\\
310118	&	17441852-2922246	&	359.4775	&	0.0158	&	 2017 May 30 	&	K,L	&		\\
314042	&	\nodata	&	359.2046	&	-0.1584	&	 2011 Jul 20 	&	K	&		\\
326277	&	17442488-2923239	&	359.4755	&	-0.0125	&	 2017 Jul 24 	&	K,L	&		\\
348392	&	\nodata	&	359.4401	&	-0.0706	&	 2009 May 15 	&	K,L	&		\\
359581	&	17443771-2736440	&	359.6667	&	0.0501	&	 2011 Jul 20 	&	K	&		\\
360444	&	17443801-2910389	&	359.6817	&	0.0579	&	 2011 Jul 19 	&	K	&		\\
362952	&	17443900-2921257	&	359.5304	&	-0.0390	&	 2011 Jul 20 	&	K	&		\\
384976	&	17444745-2903494	&	359.7966	&	0.0881	&	 2011 Jul 18 	&	K	&		\\
386185	&	17444793-2926024	&	359.4818	&	-0.1068	&	 2011 Jul 18 	&	K	&		\\
387011	&	17444824-2926055	&	359.4816	&	-0.1082	&	 2011 Jul 18 	&	K	&		\\
388790	&	17444894-2923423	&	359.5167	&	-0.0896	&	 2009 May 15 	&	K,L	&		\\
390573	&	17444960-2916489	&	359.6161	&	-0.0317	&	 2017 Jul 24 	&	K,L	&		\\
404312	&	\nodata	&	359.6630	&	-0.0255	&	 2009 May 14 	&	K,L	&		\\
405235	&	\nodata	&	359.6435	&	-0.0389	&	 2009 May 14 	&	K,L	&		\\
416141	&	17445941-2851406	&	359.9920	&	0.1567	&	 2011 Jul 18 	&	K	&		\\
423755	&	17450228-2903317	&	359.8289	&	0.0447	&	 2017 Jul 24 	&	K,L	&		\\
429933	&	17450462-2846394	&	0.0732	&	0.1842	&	 2017 Jul 23 	&	K,L	&		\\
430378	&	17450477-2905484	&	359.8013	&	0.0171	&	 2017 Jun 21 	&	K,L	&		\\
442291	&	\nodata	&	359.7697	&	-0.0215	&	 2011 Jul 18 	&	K	&		\\
452528	&	17451311-2914063	&	359.6993	&	-0.0810	&	 2017 Jul 24 	&	K,L	&		\\
453432	&	17451348-2905266	&	359.8230	&	-0.0068	&	 2017 May 28 	&	K,L	&		\\
460967	&	17451633-2909423	&	359.7678	&	-0.0527	&	 2011 Jul 18 	&	K	&		\\
463326	&	17451722-2847316	&	0.0848	&	0.1374	&	 2017 Jun 23 	&	K,L	&		\\
463777	&	17451737-2904245	&	359.8451	&	-0.0099	&	 2017 May 28 	&	K,L	&		\\
479679	&	17452327-2903309	&	359.8690	&	-0.0205	&	 2017 May 29 	&	K,L	&		\\
493484	&	17452842-2906451	&	359.8327	&	-0.0646	&	 2011 Jul 18 	&	K	&		\\
496117	&	17452941-2903561	&	359.8746	&	-0.0432	&	 2011 Jul 18 	&	K	&		\\
525210	&	17454004-2900225	&	359.9455	&	-0.0454	&	 2017 Jun 22 	&	K,L	&	GCIRS~7	\\
528212	&	17454117-2900469	&	359.9420	&	-0.0523	&	 2017 Jun 22 	&	K,L	&	GCIRS~19\\
529575	&	17454167-2852355	&	0.0594	&	0.0172	&	 2017 Jun 22 	&	K,L	&		\\
533613	&	17454315-2854218	&	0.0370	&	-0.0028	&	 2017 May 30 	&	K,L	&		\\
543304	&	17454661-2843334	&	0.1973	&	0.0802	&	 2017 Jul 23 	&	K,L	&		\\
543946	&	17454683-2856189	&	0.0162	&	-0.0312	&	 2017 Jun 23 	&	K,L	&		\\
555225	&	17455095-2835211	&	0.3223	&	0.1379	&	 2017 Jun 23 	&	K,L	&		\\
572905	&	17455743-2846306	&	0.1758	&	0.0208	&	 2017 Jul 23 	&	K,L	&		\\
576537	&	17455874-2843463	&	0.2173	&	0.0405	&	 2017 Jul 23 	&	K,L	&		\\
578515	&	17455947-2855236	&	0.0533	&	-0.0626	&	 2017 May 28 	&	K,L	&		\\
594731	&	17460531-2830553	&	0.4128	&	0.1315	&	 2017 Jun 23 	&	K,L	&		\\
597431	&	17460628-2900000	&	0.0006	&	-0.1237	&	 2017 May 30 	&	K,L	&		\\
610642	&	17461112-2855408	&	0.0712	&	-0.1012	&	 2009 May 15 	&	K,L	&		\\
614547	&	17461251-2848489	&	0.1717	&	-0.0462	&	 2017 Jun 21 	&	K,L	&		\\
617820	&	17461373-2855142	&	0.0826	&	-0.1056	&	 2017 Jun 21 	&	K,L	&		\\
620363	&	17461463-2851233	&	0.1390	&	-0.0751	&	 2017 Jul 23 	&	K,L	&		\\
625822	&	17461658-2849498	&	0.1649	&	-0.0677	&	 2017 May 29 	&	K,L	&		\\
644393	&	17462329-2847063	&	0.2165	&	-0.0650	&	 2017 May 30 	&	K,L	&		\\
650139	&	17462542-2846342	&	0.2282	&	-0.0670	&	 2017 Jun 23 	&	K,L	&		\\
653270	&	17462655-2818597	&	0.6230	&	0.1683	&	 2009 May 14 	&	K,L	&		\\
659823	&	17462894-2846164	&	0.2391	&	-0.0754	&	 2017 Jun 21 	&	K,L	&		\\
670168	&	17463266-2837184	&	0.3738	&	-0.0094	&	 2017 May 30 	&	K,L	&		\\
675433	&	17463461-2828329	&	0.5023	&	0.0603	&	 2017 Jul 23 	&	K,L	&		\\
676337	&	17463493-2821398	&	0.6010	&	0.1189	&	 2017 Jul 24 	&	K,L	&		\\
683190	&	17463748-2839363	&	0.3502	&	-0.0443	&	 2017 Jun 23 	&	K,L	&		\\
687503	&	17463903-2838570	&	0.3625	&	-0.0435	&	 2017 Jun 23 	&	K,L	&		\\
691009	&	17464029-2859367	&	0.0706	&	-0.2262	&	 2017 Jul 23 	&	K,L	&		\\
694504	&	17464159-2823182	&	0.5903	&	0.0839	&	 2017 May 28 	&	K,L	&		\\
696367	&	17464229-2833261	&	0.4472	&	-0.0060	&	 2009 May 15 	&	K,L	&		\\
709031	&	17464691-2824202	&	0.5856	&	0.0583	&	 2017 Jun 23 	&	K,L	&		\\
710552	&	17464746-2834447	&	0.4384	&	-0.0335	&	 2017 Jul 23 	&	K,L	&		\\
716531	&	17464968-2836570	&	0.4111	&	-0.0595	&	 2009 May 14 	&	K,L	&		\\
719445	&	\nodata	&	0.4921	&	-0.0149	&	 2011 Jul 17 	&	K,L	&		\\
726327E	&	17465331-2832012	&	0.4883	&	-0.0281	&	 2009 May 15 	&	K,L	&		\\
726327W	&	17465331-2832012	&	0.4880	&	-0.0279	&	2011 Jul 17	&	K,L	&		\\
730947	&	\nodata	&	0.3979	&	-0.0909	&	 2011 Jul 18 	&	K	&		\\
731318	&	17465528-2822469	&	0.6237	&	0.0455	&	 2017 May 29 	&	K,L	&		\\
734229	&	17465641-2811480	&	0.7824	&	0.1369	&	 2011 Jul 19 	&	K	&		\\
735924	&	17465707-2835039	&	0.4521	&	-0.0663	&	 2011 Jul 18 	&	K	&		\\
759523	&	17470692-2837276	&	0.4366	&	-0.1178	&	 2011 Jul 18 	&	K	&		\\
770318	&	17471173-2840486	&	0.3980	&	-0.1618	&	 2011 Jul 20 	&	K	&		\\
773805	&	17471326-2833052	&	0.5110	&	-0.0999	&	 2017 May 28 	&	K,L	&		\\
776691	&	17471454-2825045	&	0.6276	&	-0.0347	&	 2017 Jun 23 	&	K,L	&		\\
776818	&	17471459-2822192	&	0.6671	&	-0.0111	&	 2017 Jul 23 	&	K,L	&		\\
777074	&	17471472-2844013	&	0.3579	&	-0.1989	&	 2017 Jun 23 	&	K,L	&		\\
811670	&	17473040-2812526	&	0.8317	&	0.0208	&	 2017 Jun 23 	&	K,L	&		\\
812378	&	17473072-2836103	&	0.5002	&	-0.1812	&	 2017 Jul 24 	&	K,L	&		\\
817031	&	17473295-2834118	&	0.5325	&	-0.1712	&	 2009 May 15 	&	K,L	&		\\
820972	&	17473492-2815294	&	0.8030	&	-0.0159	&	 2011 Jul 19 	&	K	&		\\
841874	&	17474519-2808123	&	0.9264	&	0.0146	&	 2017 Jun 22 	&	K,L	&		\\
849203	&	17474880-2803444	&	0.9970	&	0.0417	&	 2011 Jul 20 	&	K	&		\\
853347	&	17475085-2759581	&	1.0547	&	0.0678	&	 2011 Jul 19 	&	K	&		\\
881125	&	17480558-2821202	&	0.7778	&	-0.1626	&	 2011 Jul 20 	&	K	&		\\
883563	&	17480691-2823546	&	0.7436	&	-0.1889	&	 2017 Jul 23 	&	K,L	&		\\
891918	&	17481157-2803027	&	1.0502	&	-0.0240	&	 2011 Jul 19 	&	K	&		\\
902257	&	17481734-2821233	&	0.7993	&	-0.1999	&	 2017 Jun 23 	&	K,L	&		\\   
\enddata
\end{deluxetable*}

For the observing run in 2009, we selected $11$ objects in Paper~I, as listed in Table~\ref{tab:a11}. These Spitzer/IRS sources were originally selected from the point-source catalog of the Spitzer Infrared Array Camera \citep[IRAC; ][]{fazio:04} survey in the CMZ \citep{ramirez:08}, as having extremely red mid-IR colors ($[3.6]\, -\, [8.0] \geq 2.0$); see Paper~I for more details. As described in depth in \citet{an:17}, a mid-IR source SSTGC~726327 is resolved into two near-IR objects (SSTGC~726327E and SSTGC~726327W) in a high-resolution $K$-band image from the UKIRT Infrared Deep Sky Survey \citep[UKIDSS;][]{lucas:08}. We observed the fainter one (SSTGC~726327W) in 2011, which has not been included in the observing run in 2009. In the second column of Table~\ref{tab:a11}, the classifications of these sources into a YSO ($N=2$), a ``maybe'' YSO ($N=3$), or a non-YSO ($N=7$) in Paper~I are shown, based on spectroscopic signatures of youth shown in the Spitzer/IRS spectra. The rest of the columns show observed properties of these sources from this study, including the presence/absence of CO gas absorption, the relative strength of the $3\ \micron$ H$_2$O ice band (strong, moderate, and weak), and the presence/absence of the $3.535\ \micron$ absorption from CH$_3$OH ice (see below in \S~\ref{sec:result}).

\subsection{Observing Run in 2011}

\begin{figure*}
\plottwo{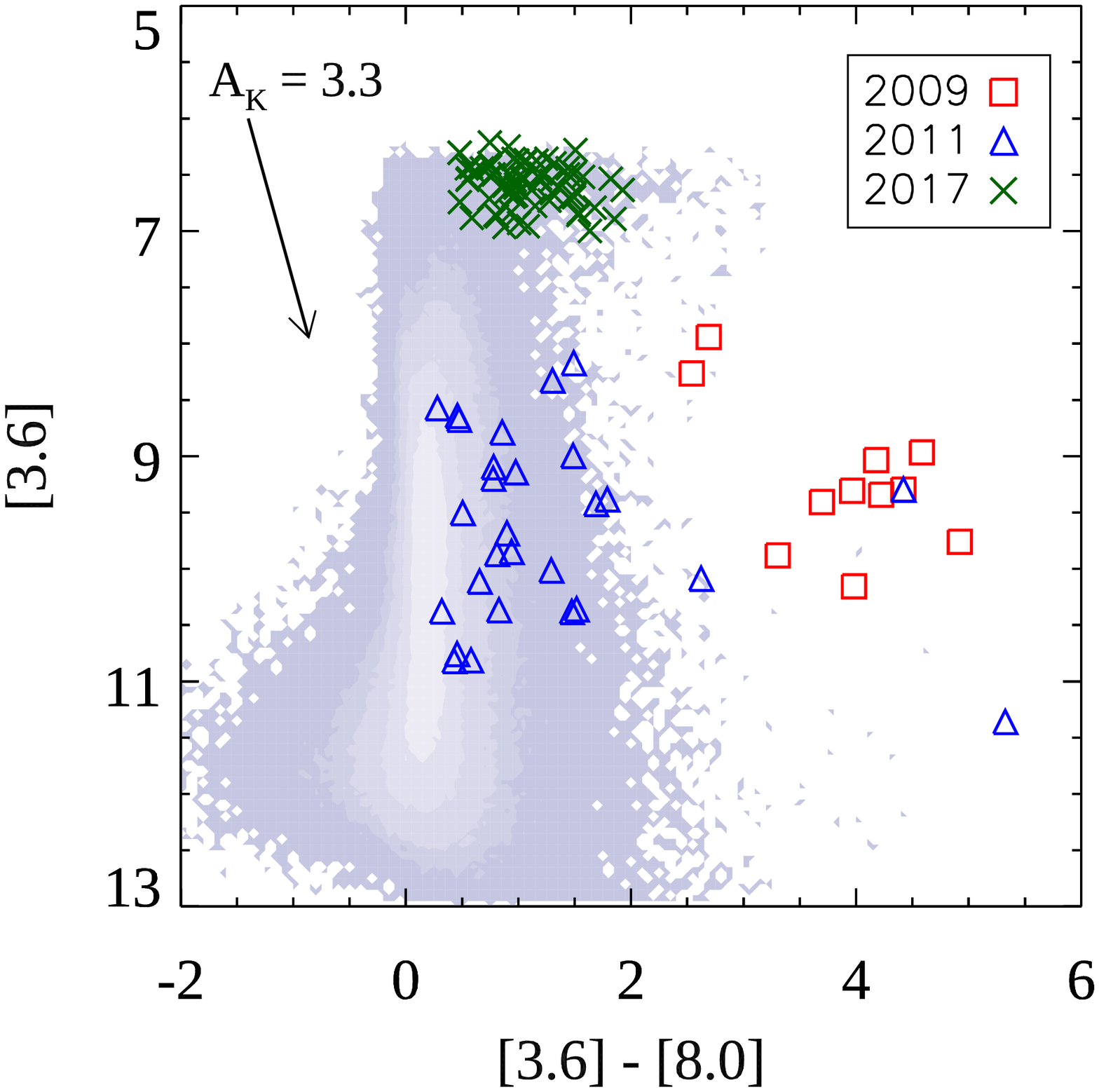}{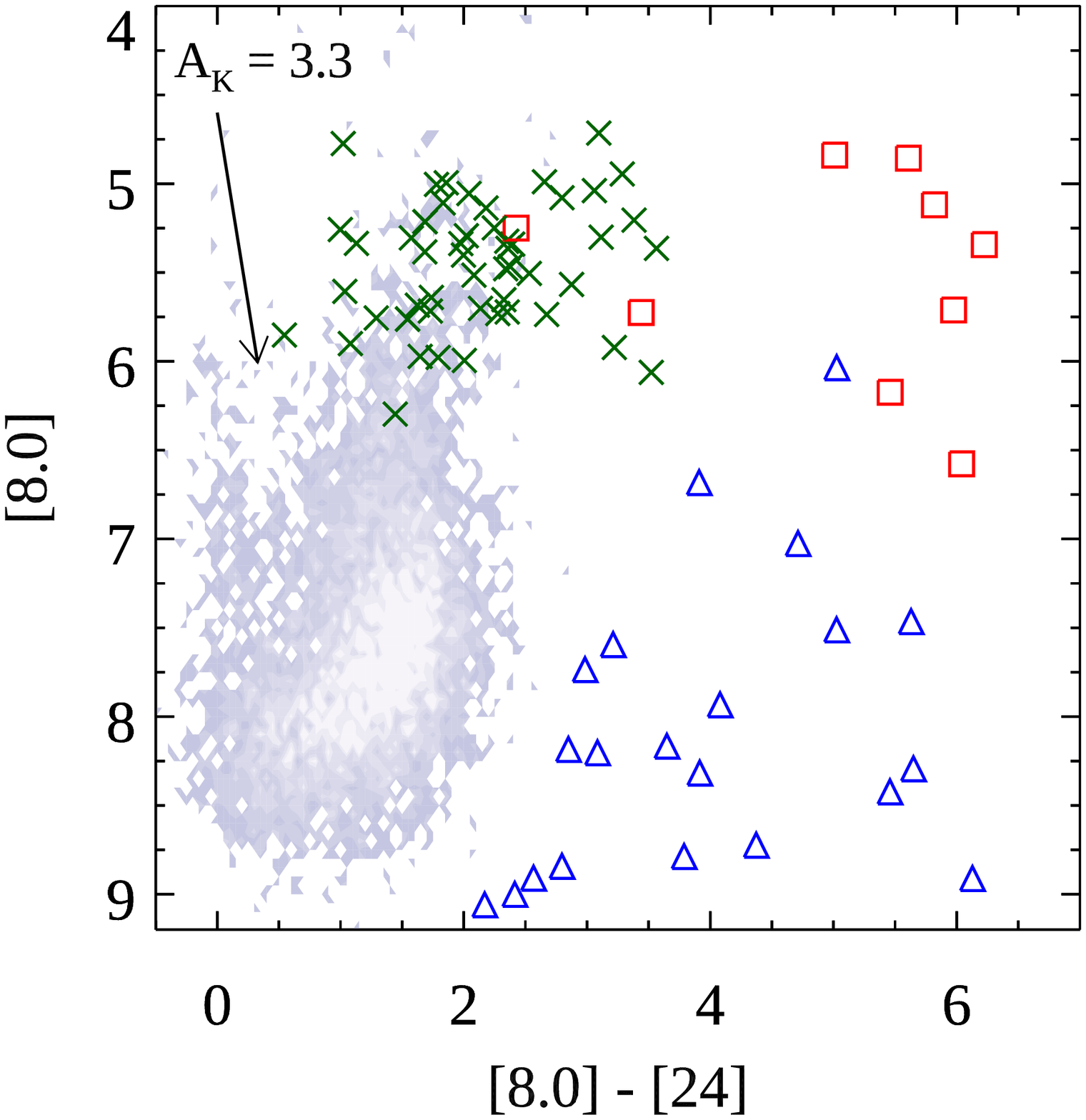}
\plottwo{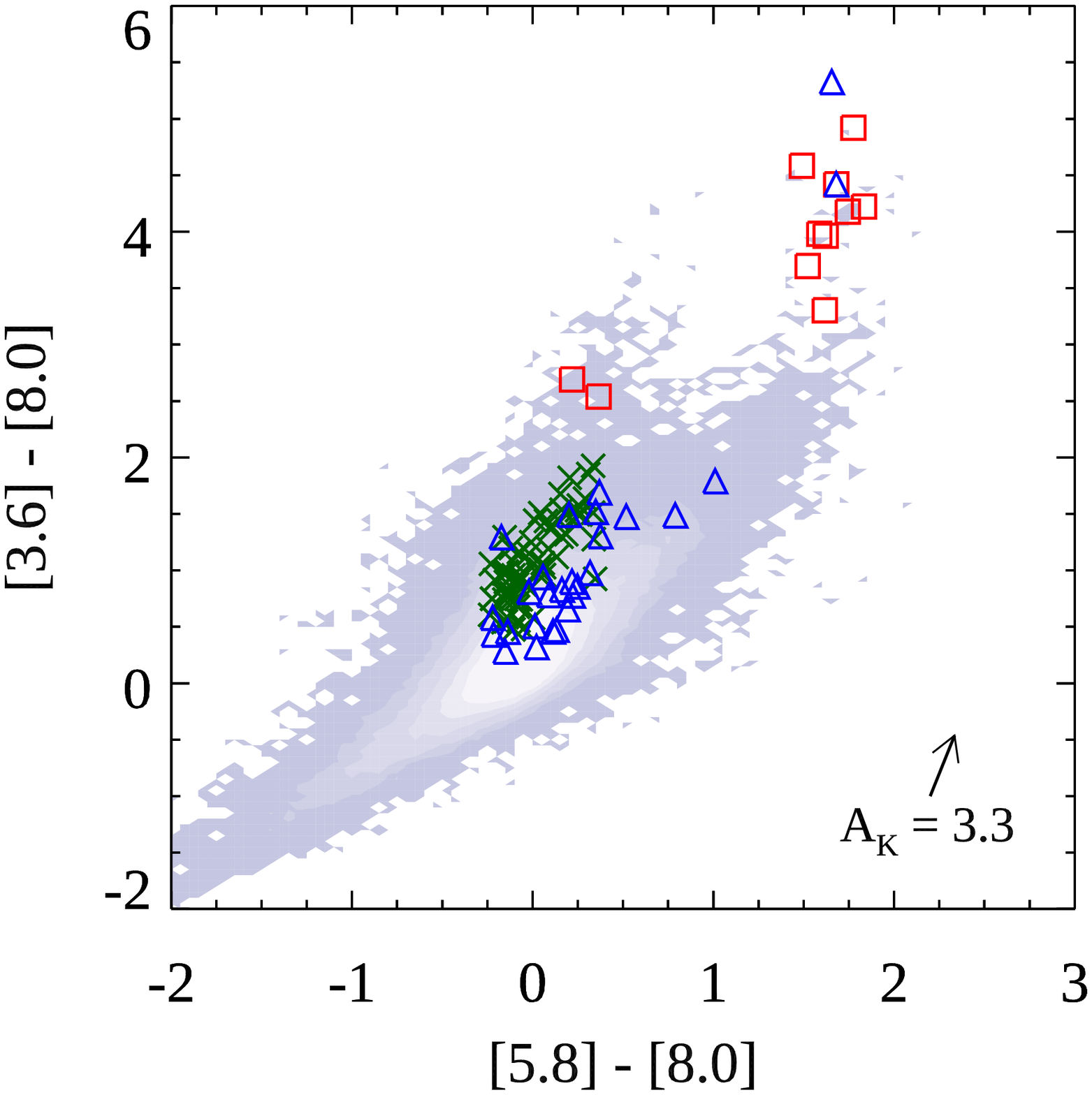}{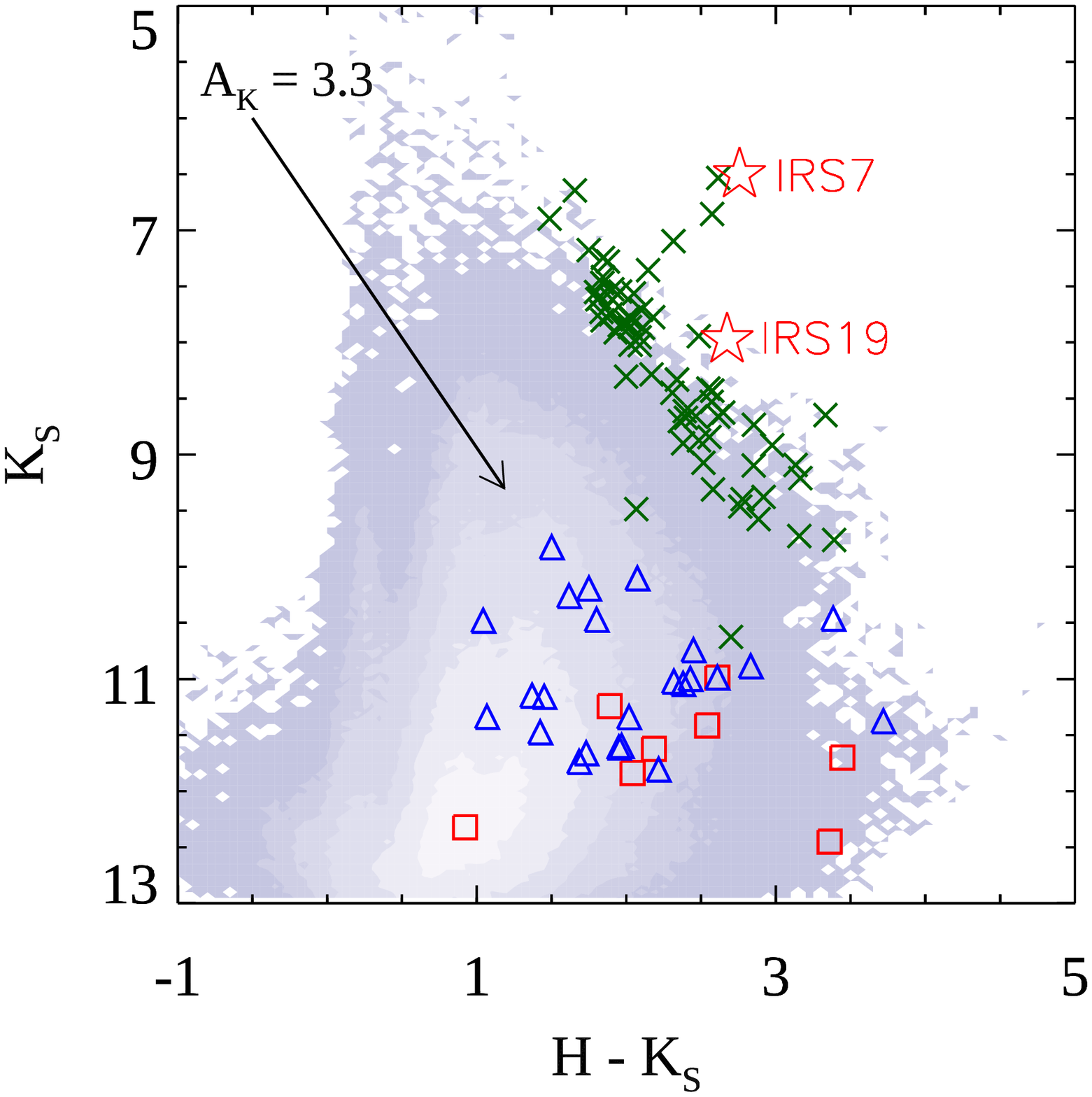}
\caption{Near- and mid-IR color-color and color-magnitude diagrams of spectroscopic samples. The sample targets are divided into three subsets: Spitzer/IRS sources from Paper~I (red boxes), photometric YSO candidates in \citet{yusefzadeh:09}, and bright IR point sources (see text) observed in 2009, 2011, and 2017, respectively. The number density distribution of stars in the CMZ is shown in gray scale, based on 2MASS, Spitzer/IRAC \citep{ramirez:08}, and Spitzer/MIPSGAL photometry \citep{gutermuth:15}. Synthetic $24\ \micron$ photometry from Paper~I is shown for the 2009 sample. GCIRS~7 and GCIRS~19 are marked by an open star symbol in the $H\, -\, K_s$ CMD, but their IRAC colors are missing due to saturation.
\label{fig:cmd}}
\end{figure*}

In 2011, we specifically targeted photometric YSO candidates in \citet{yusefzadeh:09}. They selected objects in the early stages of YSO evolution based on extremely red mid-IR colors in $[8.0]\, -\, [24]$. In total, $27$ targets were randomly selected from their study and were observed in the $K$, but not in the $L$ band.

Figure~\ref{fig:cmd} shows IR color-magnitude and color-color diagrams from the Two Micron All Sky Survey \citep[2MASS; ][]{skrutskie:06}\footnote{The photometry is available at IRSA: \dataset[doi:10.26131/IRSA2]{\doi{10.26131/IRSA2}}.} and the Spitzer/IRAC catalog in the CMZ \citep{ramirez:08}.\footnote{The photometry is available at IRSA: \dataset[doi:10.26131/IRSA505]{\doi{10.26131/IRSA505}}.} The $24\ \micron$ photometry was taken from the Spitzer/MIPSGAL survey catalogs \citep{hinz:09,gutermuth:15}.\footnote{The photometry is available at IRSA: \dataset[doi:10.26131/IRSA258]{\doi{10.26131/IRSA258}}.} Since many Spitzer/IRS sources are located on top of bright background emissions and are not found in these catalogs, owing to saturations of MIPS pixels, we used their synthetic photometry derived from Spitzer/IRS spectra in Paper~I. As shown in the $[8.0]\, -\, [24]$ versus $[8.0]$ diagram (top right panel), both Paper~I and \citet{yusefzadeh:09} objects have distinctively redder colors than background stars in the CMZ (shown by a gray Hess diagram), while their $H\, -\, K_s$ colors (bottom right panel) are indistinguishable from those of background stars. The Paper~I sample (mostly targets in 2009) also has very red $[3.6]\, -\, [8.0]$ colors (top left panel), while the majority of the \citet{yusefzadeh:09} sample (those observed in 2011) show similar $[3.6]\, -\, [8.0]$ colors with those of background sources.

\subsection{Observing Run in 2017}

In 2017, we selected $68$ objects that are brighter in $L$ than those included in the previous two observing runs. This revision in the sample selection was motivated by our earlier recognition from the previous runs that most of the spectroscopically or photometrically identified YSO candidates show strong $2.3\ \micron$ CO band-head absorptions in their $K$-band spectra, despite the fact that these absorptions are normally seen in red (super-) giants (see \S~\ref{sec:co}). This oddity has led us to expand the sample of such objects: previously known or likely (super-) giants with red mid-IR colors that are indicative of a large foreground extinction.

A subset of the targets in 2017 was taken from a list of the objects in T.\ R.\ Geballe et al. (2022, in preparation, see also \citealt{geballe:19}). They conducted a $K$-band spectroscopic survey in the CMZ to search for sources with featureless spectra, which are suitable for studying interstellar gas in this region. For the reason described above, we restricted our targets to (super-) giants identified by these authors, rejecting objects with featureless spectra or veiled CO absorptions, emission-line stars, OB stars, and Wolf-Rayet stars. The other subset came from a compilation of possible CMZ objects, constructed as part of target selections for the Apache Point Observatory Galactic Evolution Experiment of the Sloan Digital Sky Survey \citep{majewski:17}; see \citet{zasowski:13} for more details (see their Section~8.1).

All of the targets selected above have $[3.6] < 7$~mag. To ensure that they are behind extra clouds of gas, we required that they appear redder than the majority of stars observed toward the CMZ in the IR color-magnitude diagrams (see Figure~\ref{fig:cmd}), by imposing $0.46 < [3.6]\, -\, [8.0] < 2.5$, $J\, -\, K_s \geq 4.5$ and $H\, -\, K_s \geq 1.5$. There are three bright objects ($K_s < 7.1$) in the list that satisfy these $J\, -\, K_s$ and $H\, -\, K_s$ color criteria, but having no available $[3.6]\, -\, [8.0]$ colors due to saturations in the IRAC survey \citep{ramirez:08}. In addition, we observed two bright supergiants near Sgr~A$^*$, GCIRS~7 (SSTGC~525210) and GCIRS~19 (SSTGC~528212), that have been the focus of earlier studies in the literature \citep[e.g.,][]{lebofsky:82,sellgren:87,blum:96}; we used SpeX spectra of GCIRS~7 as a template of the water ice and hydrocarbon absorption band profiles (see below).

\begin{figure*}
\epsscale{1}
\plotone{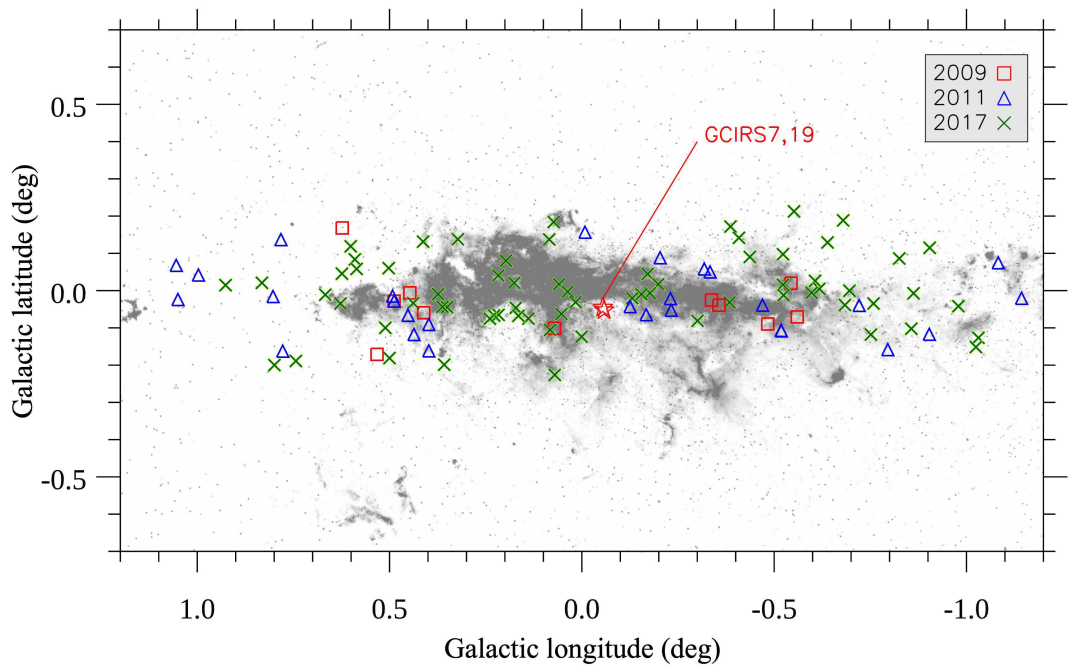}
\caption{Spatial distribution of IRTF/SpeX targets. Symbols are the same as in Figure~\ref{fig:cmd}. The background $8.0\ \micron$ image is taken from the Spitzer/IRAC broad-band survey \citep{stolovy:06}.
\label{fig:map}}
\end{figure*}

Figure~\ref{fig:map} shows a spatial distribution of our sample, overlaid on top of the Spitzer $8.0\ \micron$ IRAC image \citep{stolovy:06}. Different symbols indicate each of the three subsets of our sample, taken in 2009, 2011, and 2017, respectively. Owing to heavy visual obscuration and a large distance to the GC, none of these objects have direct distance measurements from the Gaia mission \citep{gaia_edr3}. In the absence of the three-dimensional positional information, we confined our sample to low Galactic latitudes ($|b|<0.25\arcdeg$ with a few exceptions) along the bright $8\ \micron$ structure of the CMZ clouds. Our spectroscopic targets are spread across the CMZ with no apparent grouping. Table~\ref{tab:sample} lists the final set of our spectroscopic follow-up targets.

\subsection{Data Reductions}

We followed a standard procedure of observing $K$- and/or $L'$-band spectra \citep{rayner:03} throughout the entire observing runs. For each science observation, we took telluric spectra of hot stars having spectral types between O7 to A9. For most of the cases, air mass differences and time intervals between the target and standard star observations were less than $\Delta \sec{z} = 0.2$ and $1$ hr, respectively. Because the CMZ is very crowded, we set the slit orientation of each source to minimize the contamination by nearby sources while nodding the telescope along the slit. A set of flat-field exposures and argon arcs were taken immediately before or after each standard star observation.

\begin{figure*}
\epsscale{0.9}
\plotone{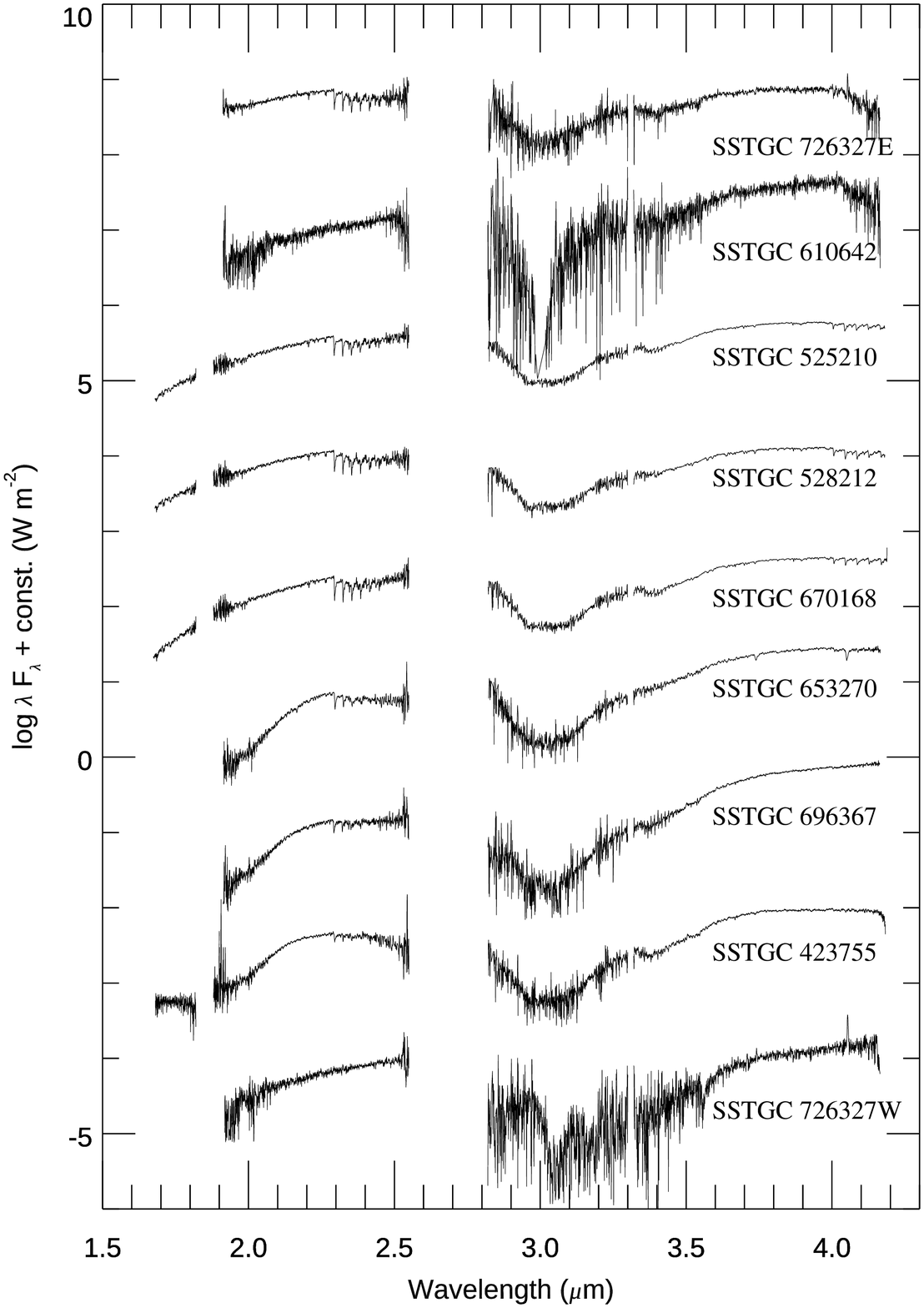}
\caption{SpeX spectra of nine objects with the strongest $3.0\ \micron$ absorption. Spectra are separated with arbitrary vertical offsets and are shown in an increasing order of $\tau_{3.0}$ from the top to the bottom.
\label{fig:spex}}
\end{figure*}

We used \texttt{Spextool} version 4.1, a spectral extraction package for SpeX \citep{cushing:04} to reduce the IRTF spectra. We combined target frames, extracted spectra using a point-source extraction mode, and combined two apertures taken from two dithering positions. For some sources, spatial profiles are wider than expected from a point source, but we proceeded with a point-source extraction with $2.0\arcsec$--$2.2\arcsec$ PSF radii anyway. Telluric absorptions were corrected following the standard procedures in \citet{vacca:03} for IRTF/SpeX spectra. Finally, after trimming noisy parts of the spectra, different orders were combined to construct a single science spectrum. Figure~\ref{fig:spex} shows SpeX spectra of nine objects with the strongest absorption at $3\ \micron$. We estimated systematic uncertainties in the flux calibration from $11$ objects that have been observed two to three times over the course of the observing runs, and found that they are at less than a $10\%$ level. 

\section{Analysis}\label{sec:analysis}

Our SpeX spectra show rich molecular absorption features that are suitable for studying the nature of our targets and the chemistry of intervening materials in the ISM and/or in an YSO envelope. We utilized a spectral library to derive the best-fitting set of a spectral type and foreground visual extinction ($A_{\rm V}$) from our $K$ and/or $L$-band observations (\S~\ref{sec:template}). The underlying continuum set by these parameters leads to accurate determination of $\tau_{3.0}$, a peak optical depth of the $3\ \micron$ band from solid H$_2$O (\S~\ref{sec:h2o}). Measurements of $\tau_{3.4}$ from the $3.4\ \micron$ hydrocarbon band (\S~\ref{sec:hc}) and $\tau_{3.54}$ from the CH$_3$OH band (\S~\ref{sec:ch3oh}) rely on a narrower, local continuum set by a polynomial equation. Table~\ref{tab:result} summarizes results from our fitting procedure, and its detailed description is provided below.

\subsection{Template Matching}\label{sec:template}

Because spectra of CMZ sources are severely attenuated by heavy foreground dust extinction, we constrained the shape of a continuum by searching for the best-fitting pair of spectral type and foreground extinction. In this template-matching procedure, we utilized a library of spectra from IRTF/SpeX \citep{rayner:09} as in our previous work \citep{an:17}, taking only those of F, G, K, and M types with full $K$- and $L$-band coverages ($218$ template spectra in total). We utilized a near-IR extinction curve in \citet{boogert:11}, which has the same shape ($A_\lambda/A_{\rm K}$) as in \citet{indebetouw:05} at $\lambda \la 3\ \micron$ (approximated as $A_\lambda \propto \lambda^{-1.8}$), but is shallower at longer wavelengths. We adopted $A_{\rm K}/A_{\rm V}=0.11$ \citep{cardelli:89}. In the following analyses, we adopted flux measurement uncertainties reported by \texttt{Spextool}, but after scaling these uncertainties to match a scatter of data with respect to a local continuum at $2.10 < \lambda < 2.25\ \micron$ (set by a second-order polynomial).

To find the best-fitting set of spectral type and foreground extinction, we computed a $\chi^2$ value for each spectral template in the library, while varying $A_{\rm V}$ and a flux normalization factor (a relative distance) to minimize its $\chi^2$ difference. In this search, we restricted our fits to $1.24\ \micron$ $< \lambda <1.32\ \micron$, $1.45\ \micron$ $< \lambda < 1.80\ \micron$, $1.93\ \micron$ $< \lambda < 2.15\ \micron$, $2.18\ \micron$ $< \lambda < 2.40\ \micron$, and $3.75\ \micron < \lambda < 4.03\ \micron$ to avoid wavelength intervals, where strong telluric absorptions or Br~$\gamma$ emission/absorption ($2.17\ \micron$) may be present. To compensate for a mild doppler shift detected in our medium-resolution spectra, we estimated a radial velocity difference using CO band heads at $2.3\ \micron$, and the above fitting procedure was iterated until satisfactory matches were found over the fitted wavelength ranges. No adjustment on the velocity difference was made if a source spectrum does not show CO absorptions (hereafter featureless spectrum).

\clearpage
\startlongtable
\begin{deluxetable*}{cccccccccc}
\tablecaption{Parameter Estimates from Spectral Fitting\label{tab:result}}
\tabletypesize{\scriptsize}
\tablehead{
   \colhead{Source ID} &
   \colhead{$A_{\rm V}$} & 
   \colhead{Spectral type} &
   \colhead{$\tau_{3.0}$} &  
   \colhead{$\tau_{3.4}$} & 
   \colhead{$\tau_{3.54}$} & 
   \colhead{$N$(H$_2$O)} & 
   \colhead{$N$(CH$_3$OH)} &
   \colhead{CO absorption} &
   \colhead{Br$\gamma$ emission} \\
   \colhead{SSTGC} & 
   \colhead{(mag)} & 
   \colhead{} & 
   \colhead{} & 
   \colhead{} & 
   \colhead{} &
   \colhead{($10^{18}$cm$^{-2}$)} & 
   \colhead{($10^{18}$cm$^{-2}$)} &
   \colhead{} &
   \colhead{}
   }
\startdata
131902 & 13.5$\pm$3.5  & K0$-$M8I$-$V  & \nodata & \nodata & \nodata & \nodata & \nodata &    &   \\
153970 & 32.7$\pm$2.5  & K1$-$M6I$-$III  & \nodata & \nodata & \nodata & \nodata & \nodata & \checkmark   &   \\
158600 & 9.5$\pm$41.6  & M7$-$M8III  & $<$0.41   & 0.10$\pm$0.01 & $<$0.01   & $<$0.67   & $<$0.03 & \checkmark   &  maybe \\
183877 & 22.9$\pm$3.0  & K1$-$M7I$-$III  & $<$0.06   & 0.04$\pm$0.01 & $<$0.01   & $<$0.11   & $<$0.05 & \checkmark   & \\
188670 & 37.2$\pm$6.4  & M5$-$M9III  & $<$0.45   & 0.11$\pm$0.04 & $<$0.01   & $<$0.74   & $<$0.05 & \checkmark  &  \checkmark \\
203189 & 37.9$\pm$15.2  & M6$-$M9III  & $<$1.90   & 0.14$\pm$0.08 & $<$0.01   & $<$3.13   & $<$0.05 & \checkmark  &  \\
207109 & 17.4$\pm$3.7  & M6$-$M7III  & $<$0.51   & $<$0.12   & $<$0.02   & $<$0.85   & $<$0.08  & \checkmark  & \\
219495 & 67.6$\pm$22.8  & K1$-$M9I$-$V & 0.29$\pm$0.10 & 0.12$\pm$0.01 & $<$0.01  & 0.48$\pm$0.16 & $<$0.03 & \checkmark  &  maybe \\
220156 & 49.4$\pm$20.4  & K0$-$M9I$-$V  & 0.43$\pm$0.36 & 0.12$\pm$0.02 & $<$0.01   & 0.71$\pm$0.60 & $<$0.02 & \checkmark  &  \\
226380 & 32.2$\pm$15.7  & M5$-$M9III  & $<$1.93   & 0.12$\pm$0.01 & $<$0.01   & $<$3.19   & $<$0.06 & \checkmark  &  \\
239331 & 32.2$\pm$5.7  & M5$-$M9III  & $<$0.48   & 0.13$\pm$0.05 & $<$0.01   & $<$0.80   & $<$0.06 & \checkmark  &  \checkmark \\
258517 & 22.1$\pm$3.3  & K0$-$M0I$-$III  & \nodata & \nodata & \nodata & \nodata & \nodata & \checkmark   &   \\
261714 & 45.9$\pm$13.3  & M2$-$M9I$-$III  & 0.84$\pm$0.46 & 0.19$\pm$0.03 & $<$0.01  & 1.39$\pm$0.75 & $<$0.05  & \checkmark  &  maybe \\
263790 & 51.1$\pm$1.4  & M7III  & 0.40$\pm$0.13 & 0.12$\pm$0.08 & $<$0.01  & 0.66$\pm$0.21 & $<$0.04 & \checkmark  &  \checkmark \\
265318 & 31.8$\pm$8.2  & M5$-$M9III$-$V  & 0.18$\pm$0.10 & 0.09$\pm$0.01 & $<$0.01   & 0.30$\pm$0.16 & $<$0.07 & \checkmark  &  \checkmark \\
268003 & 51.5$\pm$26.7  & K1$-$M9I$-$V  & $<$1.04   & 0.09$\pm$0.01  & $<$0.01   & $<$1.72   & $<$0.04 & \checkmark  & \\
272863 & 23.7$\pm$11.5  & M5$-$M9III  & 0.24$\pm$0.16 & 0.10$\pm$0.04 & $<$0.01 & 0.40$\pm$0.26 & $<$0.06 & \checkmark  &  \checkmark\\
275827 & 22.0$\pm$3.2  & K0$-$K5I$-$V  & \nodata & \nodata & \nodata & \nodata & \nodata & \checkmark   &   \\
278131 & 44.5$\pm$4.3  & M1.5$-$M7I$-$III  & 0.54$\pm$0.27 & 0.11$\pm$0.01 & $<$0.01   & 0.89$\pm$0.45 & $<$0.05 & \checkmark  &  \\
281221 & 29.6$\pm$10.7  & M3.5$-$M9III$-$V  & 0.22$\pm$0.18 & 0.11$\pm$0.03 & $<$0.01   & 0.36$\pm$0.30 & $<$0.04 & \checkmark  &  maybe \\
285458 & 42.2$\pm$7.7  & K1$-$M7.5I$-$III & 0.48$\pm$0.29 & 0.12$\pm$0.04 & $<$0.02  & 0.80$\pm$0.48 & $<$0.08 & \checkmark  &  \checkmark \\
286731 & 26.6$\pm$12.2  & K1$-$M7.5I$-$III  & $<$0.55    & 0.14$\pm$0.03 & $<$0.01   & $<$0.91    & $<$0.05 & \checkmark  &  \\
296989 & 34.6$\pm$7.4  & K0$-$M7.5I$-$III   & $<$0.01   & 0.12$\pm$0.02  & $<$0.01   & $<$0.01   & $<$0.04 & \checkmark  &  \\
298405 & 30.6$\pm$14.7  & M6$-$M9III  & $<$1.78   & 0.12$\pm$0.03  & $<$0.02   & $<$2.93    & $<$0.08 & \checkmark  &  maybe \\
300758 & 22.1$\pm$32.3  & K1$-$M9I$-$V  & $<$1.58   & $<$0.03   & $<$0.11   & $<$2.62    & $<$0.53 & maybe  &  \\
306306 & 30.6$\pm$15.1  & M5$-$M9III  & 0.37$\pm$0.23 & 0.11$\pm$0.02 & $<$0.01   & 0.61$\pm$0.38 & $<$0.04 & \checkmark  &  \checkmark \\
310118 & 56.2$\pm$11.7  & M3.5-M9I$-$III  & 0.77$\pm$0.28 & 0.23$\pm$0.04 & $<$0.01  & 1.27$\pm$0.46 & $<$0.04 & \checkmark  &  \\
314042 & 21.7$\pm$2.6  & K1.5$-$M3.5II$-$III  & \nodata & \nodata & \nodata & \nodata & \nodata & \checkmark   &   \\
326277 & 26.8$\pm$1.9  & M3.5$-$M6III  & 0.63$\pm$0.19 & $<$0.06   & $<$0.02  & 1.04$\pm$0.31 & $<$0.09 & \checkmark  &  \checkmark \\
348392 & 38.2$\pm$16.4  & K0$-$M9I$-$V & 0.76$\pm$0.57 & $<$0.05   & $<$0.08   & 1.25$\pm$0.93 & $<$0.38 &   &  \\
359581 & 31.1$\pm$3.2  & K1$-$M6I$-$III  & \nodata & \nodata & \nodata & \nodata & \nodata & \checkmark   &   \\
360444 & 25.5$\pm$2.9  & K0$-$M6I$-$III  & \nodata & \nodata & \nodata & \nodata & \nodata & \checkmark   &   \\
362952 & 33.0$\pm$4.1  & K0$-$M7I$-$III  & \nodata & \nodata & \nodata & \nodata & \nodata & \checkmark   &   \\
384976 & 21.3$\pm$2.7  & K1.5$-$M6II$-$III  & \nodata & \nodata & \nodata & \nodata & \nodata & \checkmark   &   \\
386185 & 37.4$\pm$3.7  & K1$-$M6I$-$III  & \nodata & \nodata & \nodata & \nodata & \nodata & \checkmark   &   \\
387011 & 64.0$\pm$5.5  & M3$-$M7I$-$III  & \nodata & \nodata & \nodata & \nodata & \nodata & \checkmark   &   \\
388790 & 34.5$\pm$14.0  & K0$-$M6.5I$-$V & 0.33$\pm$0.28 & 0.04$\pm$0.03  & $<$0.04  & 0.54$\pm$0.46 & $<$0.20 &   &  \\
390573 & 32.3$\pm$9.5  & M1$-$M9I$-$V   & $<$1.08   & 0.10$\pm$0.01 & $<$0.01   & $<$1.77 & $<$0.05 & \checkmark  &  \\
404312 & 36.4$\pm$26.0  & K0$-$M9I$-$V  & $<$0.45   & $<$0.39   & $<$0.05   & $<$0.73    & $<$0.24 & maybe  & \\
405235 & 53.7$\pm$24.4  & K0$-$M9I$-$V    & 0.65$\pm$0.15 & $<$0.06   & $<$0.08  & 1.07$\pm$0.24 & $<$0.39 &   & \\
416141 & 11.4$\pm$2.7  & K0$-$M6.5I$-$V  & \nodata & \nodata & \nodata & \nodata & \nodata &    &   \\
423755 & 46.5$\pm$18.5  & K4$-$M9III$-$V  & 2.67$\pm$0.64 & 0.29$\pm$0.03 & 0.10$\pm$0.01 & 4.40$\pm$1.05 & 0.47$\pm$0.02 & \checkmark  &  \\
429933 & 21.6$\pm$17.3  & M5$-$M9III  & $<$2.04   & 0.09$\pm$0.05  & $<$0.01   & $<$3.36    & $<$0.06 & \checkmark  &  maybe \\
430378 & 29.5$\pm$7.2  & M3.5$-$M8III  & 0.14$\pm$0.04 & 0.08$\pm$0.03 & $<$0.01  & 0.23$\pm$0.06 & $<$0.04 & \checkmark  &  \checkmark \\
442291 & 26.0$\pm$3.6  & K0$-$M6I$-$III  & \nodata & \nodata & \nodata & \nodata & \nodata & \checkmark   &   \\
452528 & 23.2$\pm$4.3  & M3$-$7I$-$III  & $<$0.80   & 0.07$\pm$0.04  & $<$0.02   & $<$1.32    & $<$0.09 & \checkmark  &  \\
453432 & 41.0$\pm$22.0  & K0$-$M9I$-$III  & $<$0.28 & 0.09$\pm$0.01 & $<$0.01  & $<$0.46 & $<$0.03 & \checkmark  &  \checkmark \\
460967 & 26.7$\pm$3.5  & K1$-$M6I$-$III  & \nodata & \nodata & \nodata & \nodata & \nodata & \checkmark   &   \\
463326 & 27.9$\pm$3.0  & M7III  & 0.20$\pm$0.17 & 0.05$\pm$0.01  & $<$0.01  & 0.33$\pm$0.29 & $<$0.06 & \checkmark  &  \checkmark \\
463777 & 30.5$\pm$4.2  & M3$-$M7I$-$III  & $<$0.77   & 0.08$\pm$0.04  & $<$0.02   & $<$1.27    & $<$0.08 & \checkmark  &  maybe \\
479679 & 37.6$\pm$4.0  & M6$-$M7III  & 1.03$\pm$0.19 & 0.28$\pm$0.13 & $<$0.01  & 1.70$\pm$0.32 & $<$0.06 & \checkmark  &  \\
493484 & 28.1$\pm$3.3  & M1$-$M7I$-$III  & \nodata & \nodata & \nodata & \nodata & \nodata & \checkmark   &   \\
496117 & 30.1$\pm$1.8  & M2$-$M7I$-$III  & \nodata & \nodata & \nodata & \nodata & \nodata & \checkmark   &   \\
525210 & 42.5$\pm$3.1  & K1$-$M6I$-$III  & 1.59$\pm$0.01 & 0.22$\pm$0.04 & $<$0.01  & 2.62$\pm$0.02 & $<$0.07 & \checkmark  &  \\
528212 & 40.1$\pm$3.4  & K5$-$M7I$-$III & 1.62$\pm$0.23 & 0.15$\pm$0.03 & 0.03$\pm$0.02 & 2.67$\pm$0.37 & 0.14$\pm$0.02 & \checkmark  & \\
529575 & 34.3$\pm$18.8  & M5$-$M9III  & $<$2.09   & 0.12$\pm$0.02  & $<$0.01  & $<$3.44    & $<$0.05 & \checkmark  & \\
533613 & 33.1$\pm$10.2  & K0$-$M9I$-$V  & $<$1.05   & 0.13$\pm$0.01  & $<$0.01   & $<$1.74    & $<$0.04 & \checkmark  &  maybe \\
543304 & 30.3$\pm$13.7  & M6$-$M9III  & 0.96$\pm$0.61 & 0.18$\pm$0.03 & $<$0.02  & 1.58$\pm$1.01 & $<$0.10 & \checkmark  & \\
543946 & 28.7$\pm$5.3  & K1$-$M7I$-$III  & $<$0.48   & 0.09$\pm$0.03 & $<$0.02   & $<$0.78    & $<$0.08 & \checkmark  & \\
555225 & 26.1$\pm$2.9  & M7III  & $<$0.52   & 0.10$\pm$0.04 & $<$0.01   & $<$0.85    & $<$0.06 & \checkmark  & \\
572905 & 24.5$\pm$6.0  & K1$-$M8I$-$III  & $<$0.34   & 0.08$\pm$0.01 & $<$0.01   & $<$0.56   & $<$0.07 & \checkmark  &  \checkmark \\
576537 & 24.1$\pm$3.5  & M6.5$-$M7III  & $<$0.61   & 0.10$\pm$0.07 & $<$0.01   & $<$1.00    & $<$0.07 & \checkmark & \\
578515 & 35.6$\pm$16.7  & M5$-$M9III$-$V  & $<$0.68   & 0.11$\pm$0.02 & $<$0.01   & $<$1.12   & $<$0.05 & \checkmark  &  \\
594731 & 26.6$\pm$17.4  & K2.5$-$M9II$-$V & $<$1.31   & 0.11$\pm$0.01 & $<$0.01   & $<$2.15   & $<$0.05 & \checkmark  &  \checkmark \\
597431 & 28.7$\pm$20.6  & M6$-$M9III  & $<$2.20   & 0.28$\pm$0.05 & $<$0.01   & $<$3.63   & $<$0.06 & \checkmark  &  \\
610642 & 51.9$\pm$27.6  & K0$-$M9I$-$V & 1.46$\pm$0.29 & $<$0.18   & $<$0.07   & 2.41$\pm$0.49 & $<$0.32 &   & \\
614547 & 46.9$\pm$7.0  & M5$-$M9III  & 0.74$\pm$0.15 & 0.13$\pm$0.08 & $<$0.01  & 1.21$\pm$0.25 & $<$0.06 & \checkmark  & \\
617820 & 43.5$\pm$9.7  & M5$-$M9III  & 0.33$\pm$0.32 & 0.14$\pm$0.03 & $<$0.01  & 0.54$\pm$0.52 & $<$0.04 & \checkmark  & \\
620363 & 23.6$\pm$4.2  & M6$-$M7III  & $<$0.56   & 0.07$\pm$0.01  & $<$0.02   & $<$0.92    & $<$0.08 & \checkmark  &  maybe \\
625822 & 25.3$\pm$2.8  & K1$-$M7I$-$III  & 0.39$\pm$0.19 & 0.08$\pm$0.03 & $<$0.03  & 0.64$\pm$0.31 & $<$0.08 & \checkmark  & \\
644393 & 38.0$\pm$9.1  & M5$-$M9III  & 0.80$\pm$0.29 & 0.19$\pm$0.05 & $<$0.01   & 1.32$\pm$0.48 & $<$0.06 & \checkmark  & \\
650139 & 23.3$\pm$3.5  & M6.5$-$M7III  & $<$0.58   & 0.05$\pm$0.02 & $<$0.01  & $<$0.96   & $<$0.07 & \checkmark  &  maybe \\
653270 & 53.6$\pm$17.7  & M5$-$M9III  & 1.82$\pm$0.64 & 0.19$\pm$0.13  & 0.04$\pm$0.01 & 3.01$\pm$1.05 & 0.18$\pm$0.02 & \checkmark  & \\
659823 & 51.1$\pm$8.9  & M3.5$-$M7III  & 0.61$\pm$0.29 & 0.13$\pm$0.02  & $<$0.01  & 1.00$\pm$0.48 & $<$0.04 & \checkmark &  maybe \\
670168 & 49.0$\pm$5.2  & K1$-$M7I$-$III & 1.65$\pm$0.28 & 0.28$\pm$0.01  & 0.05$\pm$0.03 & 2.72$\pm$0.46 & 0.22$\pm$0.02 & \checkmark  & \\
675433 & 24.6$\pm$8.2  & K0$-$M8I$-$III  & 0.29$\pm$0.18 & 0.08$\pm$0.01 & $<$0.01  & 0.48$\pm$0.30 & $<$0.04 & \checkmark  &  \\
676337 & 31.7$\pm$18.1  & K1$-$M9I$-$V  & $<$1.14 & 0.14$\pm$0.01 & $<$0.01  & $<$1.88 & $<$0.04 & \checkmark  &  \\
683190 & 31.0$\pm$4.4  & M1.5$-$M7I$-$III  & 0.38$\pm$0.20 & 0.10$\pm$0.02 & $<$0.01  & 0.63$\pm$0.32 & $<$0.06 & \checkmark  &  \\
687503 & 26.7$\pm$3.1  & M7III  & 0.22$\pm$0.18   & 0.08$\pm$0.01  & $<$0.02  & 0.36$\pm$0.29   & $<$0.08 & \checkmark  &  \\
691009 & 32.1$\pm$6.1  & K1$-$M8I$-$III  & $<$0.73   & 0.08$\pm$0.05  & $<$0.01   & $<$1.21   & $<$0.04 & \checkmark  &  \\
694504 & 41.5$\pm$17.7  & M5$-$M9III  & 0.84$\pm$0.34 & 0.12$\pm$0.06 & $<$0.02  & 1.39$\pm$0.56 & $<$0.05 & \checkmark  &  \\
696367 & 51.0$\pm$25.3  & K1.5$-$M9III$-$V & 2.38$\pm$0.91 & 0.24$\pm$0.02 & 0.05$\pm$0.03 & 3.93$\pm$1.51 & 0.25$\pm$0.04 & \checkmark  &  \\
709031 & 26.2$\pm$4.3  & M3$-$M7I$-$III  & $<$0.79   & 0.07$\pm$0.03  & $<$0.01   & $<$1.30   & $<$0.07 & \checkmark  &  maybe \\
710552 & 26.5$\pm$4.1  & M3$-$M7I$-$III  & $<$0.76   & $<$0.21   & $<$0.01   & $<$1.26   & $<$0.06 & \checkmark  &  maybe \\
716531 & 31.8$\pm$16.7  & K0$-$M9I$-$V & $<$1.07   & $<$0.30    & $<$0.05   & $<$1.76   & $<$0.24 &   &  \\
719445 & 41.6$\pm$25.7  & K0$-$M9I$-$V  & \nodata & \nodata & \nodata & \nodata & \nodata &    &   \\
726327E & 33.7$\pm$4.9  & K1$-$M7I$-$III  & 1.45$\pm$0.24 & 0.25$\pm$0.06 & 0.09$\pm$0.08 & 2.39$\pm$0.39 & 0.42$\pm$0.18 & \checkmark  &  \checkmark \\
726327W & 66.8$\pm$10.2  & K0$-$M6.5III$-$V   & 3.87$\pm$0.36 & $<$0.36   & $<$0.06   & 6.39$\pm$0.60 & $<$0.26 &   &  maybe \\
730947 & 26.8$\pm$4.2  & K0$-$M0I$-$V  & \nodata & \nodata & \nodata & \nodata & \nodata & \checkmark   &   \\
731318 & 27.3$\pm$19.5  & M5$-$M9III  & $<$2.25   & 0.10$\pm$0.04 & $<$0.01   & $<$3.71    & $<$0.05 & \checkmark  &  maybe \\
734229 & 43.3$\pm$38.5  & K1$-$M9I$-$V  & \nodata & \nodata & \nodata & \nodata & \nodata & \checkmark &   \\
735924 & 21.7$\pm$3.3  & K1$-$M6I$-$III  & \nodata & \nodata & \nodata & \nodata & \nodata & \checkmark   &   \\
759523 & 47.8$\pm$5.9  & M7$-$M8III  & \nodata & \nodata & \nodata & \nodata & \nodata & \checkmark   &   \\
770318 & 33.6$\pm$3.8  & K1$-$M7I$-$III  & \nodata & \nodata & \nodata & \nodata & \nodata & \checkmark   &   \\
773805 & 46.7$\pm$12.5  & K2.5$-$M9II$-$V  & 1.09$\pm$0.21 & 0.21$\pm$0.02  & $<$0.01   & 1.80$\pm$0.35 & $<$0.06 & \checkmark  &  \\
776691 & 29.4$\pm$3.2  & M7III  & 0.53$\pm$0.19 & $<$0.09   & $<$0.03   & 0.87$\pm$0.31 & $<$0.08 & \checkmark  &  \\
776818 & 24.7$\pm$18.3  & M5$-$M9III  & 0.83$\pm$0.70 & 0.16$\pm$0.03 & $<$0.01  & 1.37$\pm$1.15 & $<$0.07 & \checkmark  &  \\
777074 & 23.9$\pm$3.2  & M7III  & 0.46$\pm$0.18 & $<$0.09   & $<$0.01   & 0.76$\pm$0.31 & $<$0.05 & \checkmark  &  \\
811670 & 26.6$\pm$9.2  & M3.5$-$M8I$-$III  & 0.26$\pm$0.04 & 0.10$\pm$0.02 & $<$0.02   & 0.43$\pm$0.06 & $<$0.08 & \checkmark  &  \\
812378 & 34.5$\pm$14.3  & M6$-$M9III  & $<$1.85   & 0.19$\pm$0.07 & $<$0.01   & $<$3.05   & $<$0.06 & \checkmark  &  \\
817031 & 30.7$\pm$15.0  & K0$-$M9I$-$V & 0.70$\pm$0.45 & $<$0.33   & $<$0.09   & 1.16$\pm$0.74 & $<$0.40 & \checkmark  &  \\
820972 & 35.8$\pm$3.8  & K1$-$M6I$-$III  & \nodata & \nodata & \nodata & \nodata & \nodata & \checkmark   &   \\
841874 & 34.3$\pm$18.2  & K1$-$M9I$-$V  & 0.69$\pm$0.45 & 0.17$\pm$0.04 & $<$0.01  & 1.14$\pm$0.74 & $<$0.04 & \checkmark  &  \\
849203 & 47.0$\pm$2.6  & K0$-$M6I$-$III  & \nodata & \nodata & \nodata & \nodata & \nodata & \checkmark   &   \\
853347 & 35.0$\pm$3.6  & K1$-$M6I$-$III  & \nodata & \nodata & \nodata & \nodata & \nodata & \checkmark   &   \\
881125 & 24.8$\pm$3.7  & K0$-$M2I$-$V  & \nodata & \nodata & \nodata & \nodata & \nodata & \checkmark   &   \\
883563 & 29.3$\pm$5.6  & M5$-$M9III  & 0.25$\pm$0.17 & 0.11$\pm$0.04 & $<$0.01   & 0.41$\pm$0.28 & $<$0.06 & \checkmark  &  maybe \\
891918 & 42.3$\pm$4.0  & K1$-$M6I$-$III  & \nodata & \nodata & \nodata & \nodata & \nodata & \checkmark   &   \\
902257 & 35.7$\pm$11.8  & M3$-$M7I$-$III  & $<$1.28   & 0.10$\pm$0.02 & $<$0.01   & $<$2.12    & $<$0.05 & \checkmark  &  maybe \\
\enddata
\end{deluxetable*}

\begin{figure*}
\epsscale{0.7}
\plotone{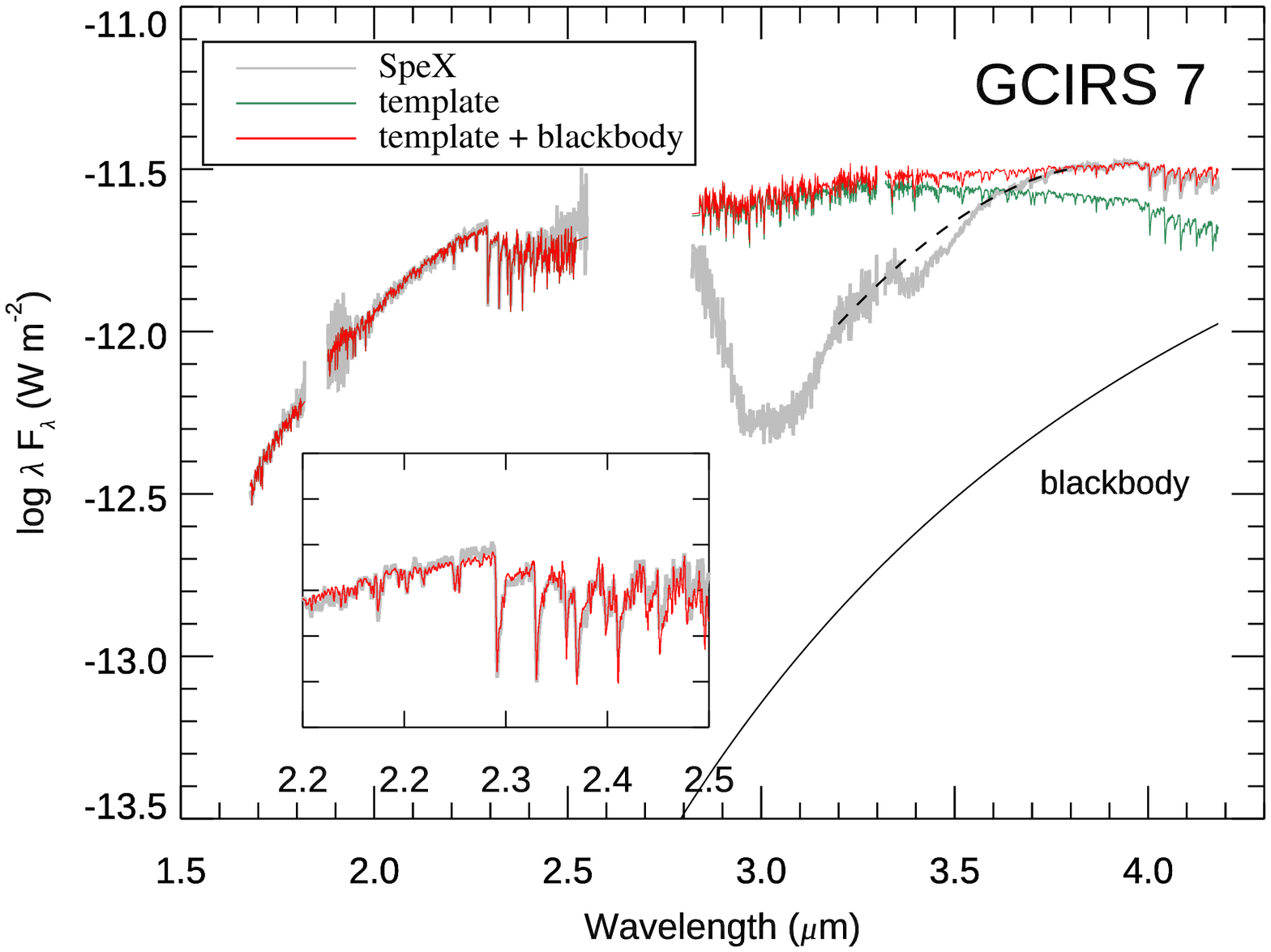}
\epsscale{0.9}
\plottwo{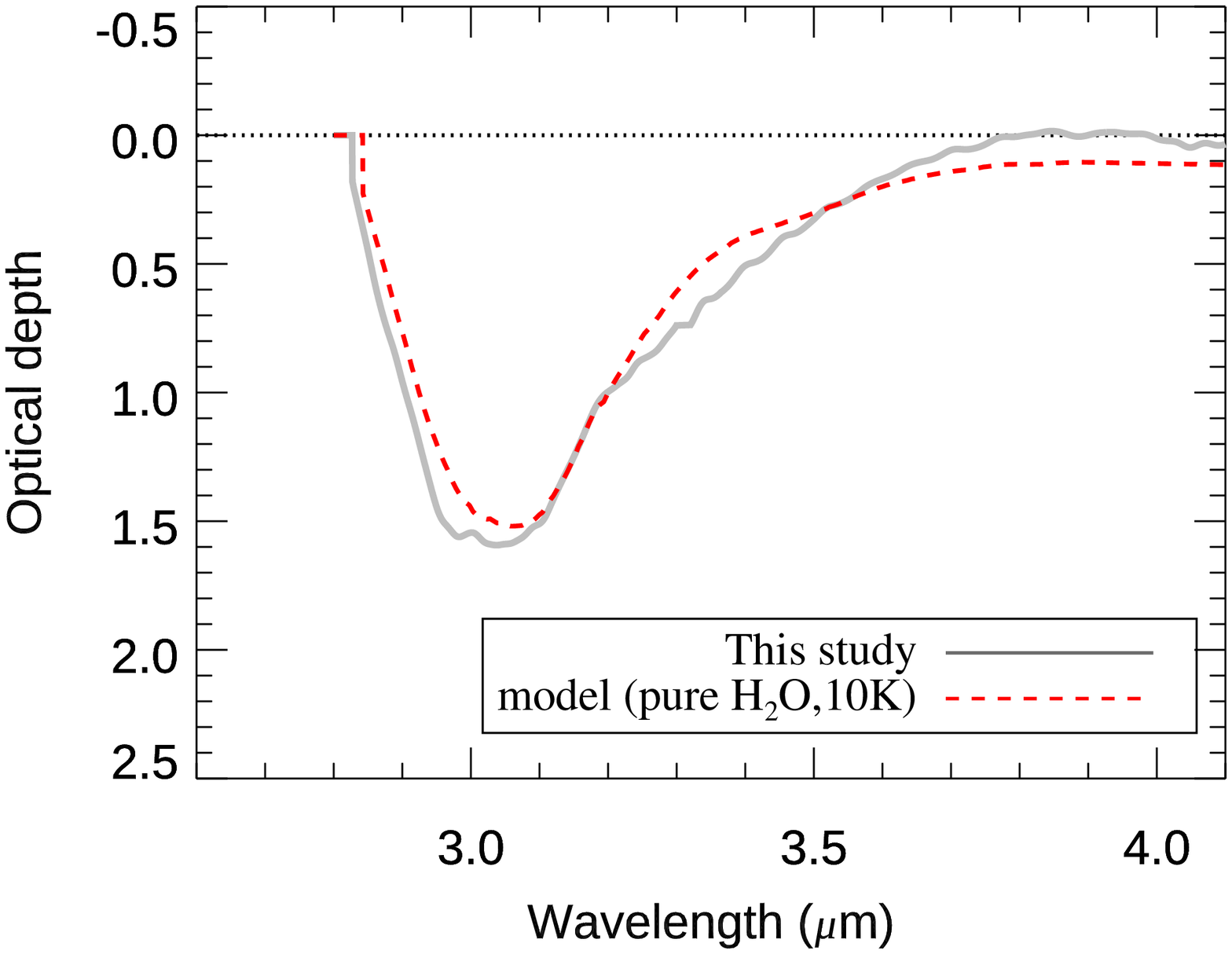}{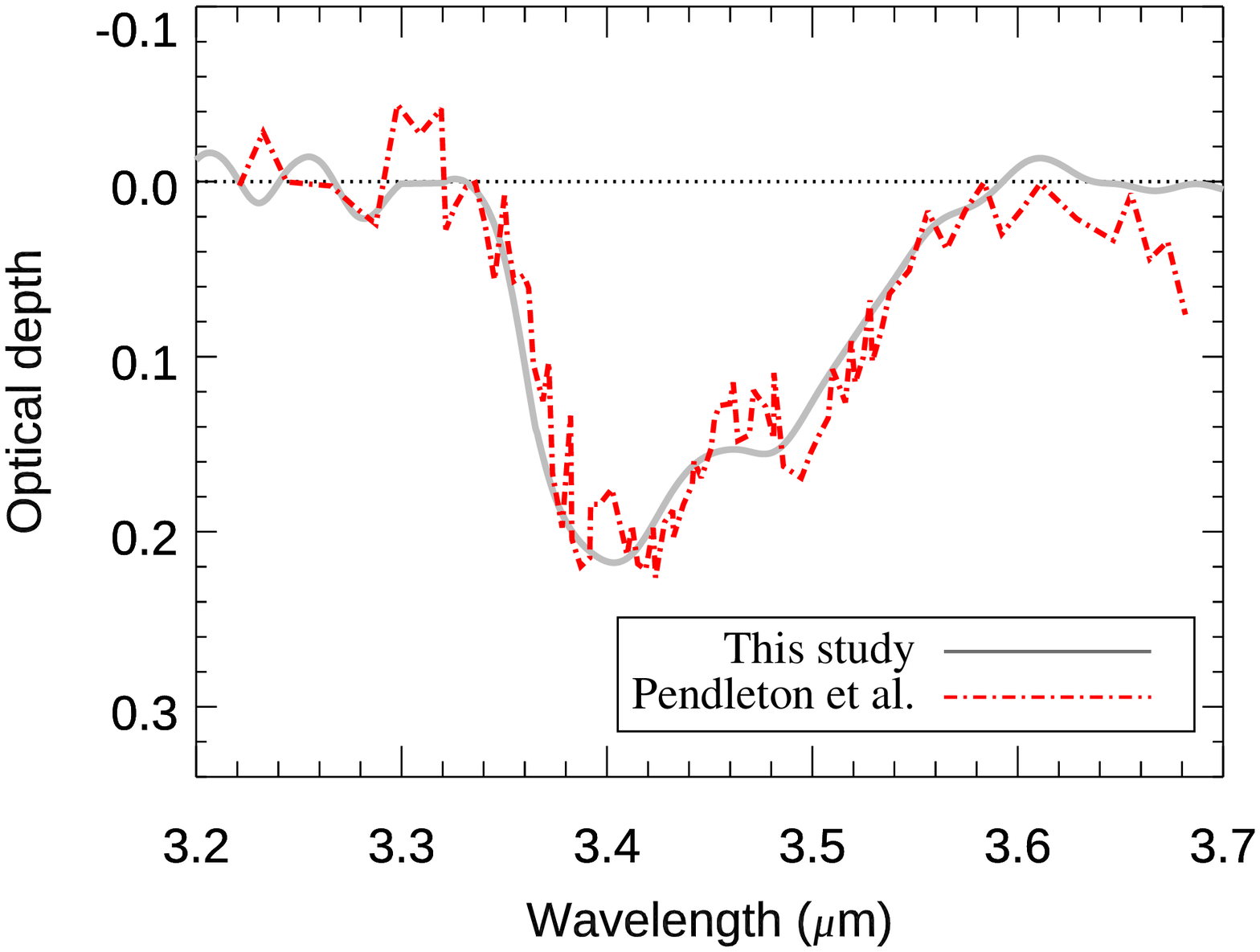}
\caption{Spectral fitting for GCIRS~7 (SSTGC~525210; gray line). Top: the green line is an IRTF/SpeX template \citep[M5.5III;][]{rayner:09} with $A_{\rm K}=4.7$, which shows the best match to our observed spectra at $1.24\ \micron$ $< \lambda <1.32\ \micron$, $1.45\ \micron$ $< \lambda < 1.80\ \micron$, $1.93\ \micron$ $< \lambda < 2.15\ \micron$, and $2.18\ \micron$ $< \lambda < 2.40\ \micron$. The red line is a sum of contributions from the green line and the best-fitting blackbody curve ($400$~K) to the residual at $3.75\ \micron$ $< \lambda < 4.03\ \micron$. The inset figure shows a narrow wavelength interval near $\sim2.3\ \micron$ CO absorption. Bottom left: an optical depth spectrum from the best-fitting IRTF template spectrum (red line in the top panel). The hydrocarbon absorption band centered at $\sim3.4\ \micron$ is masked using a third-order polynomial at $3.2\ \micron$ $< \lambda < 3.8\ \micron$ (black dashed line in the top panel). For comparison, the red dashed line shows a laboratory spectrum of pure H$_2$O ice on $0.8\ \micron$ ice spheres at $10$~K \citep{hudgins:93}, scaled to match the GCIRS~7 spectrum. Bottom right: an optical depth spectrum derived from a local continuum (black dashed line in the top panel). For comparison, the red dotted-dashed line shows a spectrum of GCIRS~7 in \citet{pendleton:94}. The spectra in the lower two panels are smoothed in order to be used as a template for the H$_2$O ice and hydrocarbon absorption bands.
\label{fig:fitting}}
\end{figure*}

Figure~\ref{fig:fitting} shows a result from a template matching for GCIRS~7 (SSTGC~525210), an M2 red supergiant near Sgr~A$^*$ \citep{lebofsky:82,sellgren:87}. As shown in this example, our fitting produces a reasonable match to the $K$-band region, including the shape and the strength of CO absorption bands. However, we found, based on fits to other spectra, that the best-fitting template spectrum tends to exhibit systematically weaker emissions at $\ga3.7\ \micron$. If a template is forced to match the $L$-band region ($3.75\ \micron$ $< \lambda < 4.03\ \micron$), a resultant fit becomes progressively worse at shorter wavelengths, especially in $K$.

The observed discrepancy in the long-wavelength region can be caused by systematic errors in the flux calibration and/or poor background subtraction of SpeX spectra, because the CMZ exhibits bright and complex structures of diffuse mid-IR emissions \citep{stolovy:06}. Background emissions over a few arc-second scale or larger should be effectively subtracted by our telescope nodding. Nonetheless, compact (an angular scale of a few arc-seconds or less) emissions are still difficult to remove, and therefore can contribute to excess emissions in the long wavelengths. On the other hand, it may also be that the excess mid-IR emission has an astrophysical origin. As presented in this work, there is evidence indicating that some of our targets are located behind a localized warm dust cloud. In support of this view, we find that a warm blackbody spectrum ($300$--$400$~K) is required to fit an overall shape of the Spitzer/IRS spectra in Paper~I, in addition to colder ($\sim100$~K) dust component(s) (see below in \S~\ref{sec:726327}).

To compensate for the observed offset at $\ga3.7\ \micron$, we added a single blackbody spectrum (black solid line) as an extra component in the modeling. The resultant fit for GCIRS~7 is shown by a red solid line in Figure~\ref{fig:fitting}, which clearly exhibits an improved fit over the entire wavelength range. However, because a blackbody curve was constrained over a relatively small wavelength interval ($3.75\ \micron$ $< \lambda < 4.03\ \micron$), we could only have a meaningful constraint on the flux normalization, but not its shape parameter; therefore, our derived blackbody temperatures remain close to our best initial guess ($400$~K). Nonetheless, since the excess emission at $\lambda\ga3.7\ \micron$ is found in the long-wavelength edge of the $3\ \micron$ absorption band, adding an extra component to the model has a little impact on the derivation of $\tau_{3.0}$ as well as abundances of other molecular species in this study.

Except for the \citet{yusefzadeh:09} sample, we have both $K$- and $L$-band coverages ($82$ objects including SSTGC~726327W), based on which we derived $A_{\rm V}$ and spectral type from a minimum $\chi^2$ value of the fit ($\chi_{\rm min}^2$). For the $27$ objects in \citeauthor{yusefzadeh:09}, we attempted to obtain these parameters from $K$-band spectra only. To estimate confidence intervals of our parameter estimates, we computed $1\sigma$ random uncertainties from $\Delta \chi^2 < 4.72$ with respect to the $\chi^2_{\rm min}$, after multiplying a $\Delta \chi^2$ distribution by $\nu/\chi_{\rm min}^2$, where $\nu$ is a degree of freedom in our parameter search. Here, we assumed four degrees of freedom (spectral type, extinction, flux normalization, and a normalization of a blackbody curve), because our fitting is essentially independent of a blackbody temperature (see above). For $A_{\rm V}$, we added this in quadrature to a difference in $A_{\rm V}$ obtained from the case, where the spectral type was fixed to M5.5III (the best-fitting spectral type for GCIRS~7). Nonetheless, the sizes of these uncertainties are comparable to each other, possibly indicating overestimation of a total uncertainty in $A_{\rm V}$ by a factor of $\sim\sqrt{2}$. In Table~\ref{tab:result}, the best-fitting set of extinction and spectral type, and their uncertainties are listed in the second and third columns, respectively. We note that a spectral type (in particular, a luminosity class) is a rather poorly constrained quantity in our template matching, owing to a relatively narrow range of wavelength in our SpeX spectra.

In addition, we explored other IR extinction curves \citep{chiar:06,fritz:11} to evaluate the impact of different extinction laws on our parameter estimates. The extinction curve in \citet{chiar:06} is essentially based on a combination of \citet{lutz:99} and \citet{indebetouw:05}, and provides a slightly steeper slope in near IR than our base case. The extinction curve in \citet{fritz:11} has a power-law index $\alpha = -2.1$, and provides an even steeper slope in near IR, similarly to other recent determinations toward the GC \citep{nishiyama:09,schodel:10}. We repeated the above template matching, assuming each of these extinction curves, and found that the mean differences from those in Table~\ref{tab:result} are $\langle \Delta A_{\rm V} \rangle = 2.4\pm0.2$~mag and $11.3\pm0.5$~mag with respect to \citet{chiar:06} and \citet{fritz:11}, respectively, in the sense of larger $A_{\rm V}$ from our default case. Although $\sim10\%$--$30\%$ differences are quite significant, they are systematic in nature. Since only the {\it relative} difference in $A_{\rm V}$ is relevant in this work (\S~\ref{sec:av}), they are not included in our error budget in Table~\ref{tab:result}.

\subsection{Solid H$_2$O}\label{sec:h2o}

In this study, we attributed the $3\ \micron$ absorption band entirely to H$_2$O ice, since H$_2$O ice is the most abundant species among other carriers that are responsible for the O--H stretching mode, and its band strength is also larger \citep{boogert:15}. The laboratory spectrum for pure H$_2$O ice produces a single peak at $\sim3.1\ \micron$, and matches the observed spectrum of a star behind dense molecular cloud cores \citep{boogert:11} reasonably well. \citet{chiar:02} used such models to achieve a ``global'' fit to GC source spectra, suggesting that the absorption arises mainly from cold, quiescent clouds as in the Galactic disk. \citet{an:17} also adopted a laboratory spectrum of pure amorphous H$_2$O ice on $0.8\ \micron$ ice spheres at $10$~K \citep{hudgins:93} to find a reasonable fit to the spectrum of SSTGC~726327. However, the observed $3\ \micron$ absorption band often exhibits double peaks at $\sim2.95\ \micron$ and $\sim3.05\ \micron$ \citep{chiar:02}, likely originated from crystalline water ice and NH$_3$ hydrates, respectively \citep{chiar:00,boogert:15}; see the bottom left panel in Figure~\ref{fig:fitting}. In addition, the shape of the $3\ \micron$ band is sensitively affected by grain-size distributions and the degree of thermal processing of ices \citep{smith:89}. Toward the GC, $3.3\ \micron$ absorption was also attributed to polycyclic aromatic hydrocarbons \citep{chiar:02}. Therefore, varying physical and chemical conditions along each line of sight through local clouds, intervening spiral arms, CMZ clouds, and/or a YSO envelope collectively influence the observed absorption band.

Due to complex substructures of the $3\ \micron$ band, we took a purely empirical approach using a template H$_2$O ice profile, and defer a detailed modeling to future studies. For this purpose, we chose the GCIRS~7 spectrum as a template, since it has a relatively high SNR and a strong $3\ \micron$ absorption band (Figure~\ref{fig:fitting}). To extract a smooth H$_2$O ice profile, we ran a moving boxcar average, after masking the $3.4\ \micron$ absorption band complex using a third-order polynomial (fitted at $3.29$--$3.31\ \micron$ and $3.60$--$3.62\ \micron$). For the rest of the objects, we scaled the smoothed GCIRS~7 template to match data at $2.9\ \micron$ $< \lambda < 3.2\ \micron$ and $3.7\ \micron$ $< \lambda < 3.9\ \micron$.

In Table~\ref{tab:result}, the fourth column contains the optical depth at $3\ \micron$ ($\tau_{3.0}$) of the best-matching, scaled GCIRS~7 template profile. The uncertainty in $\tau_{3.0}$ represents a quadrature sum of (i) an uncertainty computed from the covariance matrix and (ii) a systematic uncertainty from varying continuum levels by different combinations of $A_{\rm V}$ and spectral type within their $1\sigma$ confidence intervals. In order to derive a column density of solid H$_2$O, we multiplied $\tau_{3.0}$ by FWHM $=330$~cm$^{-1}$ for amorphous H$_2$O ice grains, instead of directly integrating over the fitted H$_2$O template spectrum, owing to absorption features from other molecular species; effectively, we assumed that the peak of the absorption band is dominated by H$_2$O ice. We adopted a band strength of $A = 2.0 \times 10^{-16}$~cm~molecule$^{-1}$ \citep{hagen:81}. The uncertainty in $N$(H$_2$O) was directly propagated from the uncertainty in $\tau_{3.0}$.

\subsection{Hydrocarbons}\label{sec:hc}

The bottom right panel in Figure~\ref{fig:fitting} shows an optical depth spectrum of GCIRS~7 with respect to a local continuum set by a third-order polynomial, fitted to $3.20$--$3.35\ \micron$ and $3.60$--$3.80\ \micron$ (the black dashed line in the top panel). This absorption band complex at $3.4\ \micron$ is ascribed to stretching vibration of C--H of methyl (--CH$_3$) and methylene (--CH$_2$--) groups of aliphatic hydrocarbons. For comparison, an optical depth spectrum from \citet{pendleton:94}\footnote{The data were originally taken from \citet{sandford:91}.} is displayed by a red dotted-dashed line, which shows substructures with narrow absorptions at $3.385$, $3.420$, and $3.485\ \micron$ \citep{sandford:91,pendleton:94}. By design, our {\it smoothed} spectrum does not clearly exhibit such substructures, but we note that its overall shape remains in good agreement with theirs.

As for the H$_2$O ice band, we constructed a template spectrum for the $3.4\ \micron$ complex from GCIRS~7, and scaled it to match spectra over $3.35$--$3.45\ \micron$ in order to derive a peak optical depth at $3.4\ \micron$ ($\tau_{3.4}$). For internal consistency, we set a local continuum using the same polynomial order and wavelength ranges as for GCIRS~7. In most cases, we found that the shape of the observed absorption band is similar to that of GCIRS~7, while their overall strength varies from one to the other. The only exception is SSTGC~479679, which shows an exceptionally strong absorption in one of the narrow absorption features at $3.385\ \micron$.

Table~\ref{tab:result} lists $\tau_{3.4}$ in the fifth column, only for those with $L$-band measurements. We estimated the uncertainty in $\tau_{3.4}$ as a quadrature sum of (i) a fitting uncertainty and (ii) a systematic component from a continuum setting. For the latter, we employed an alternative local continuum using a first-order polynomial, fitted over $3.29$--$3.31\ \micron$ and $3.60$--$3.62\ \micron$, and took its difference from our base case as an effective $1\sigma$ uncertainty.

\subsection{Solid CH$_3$OH}\label{sec:ch3oh}

\begin{figure}
\epsscale{1.2}
\plotone{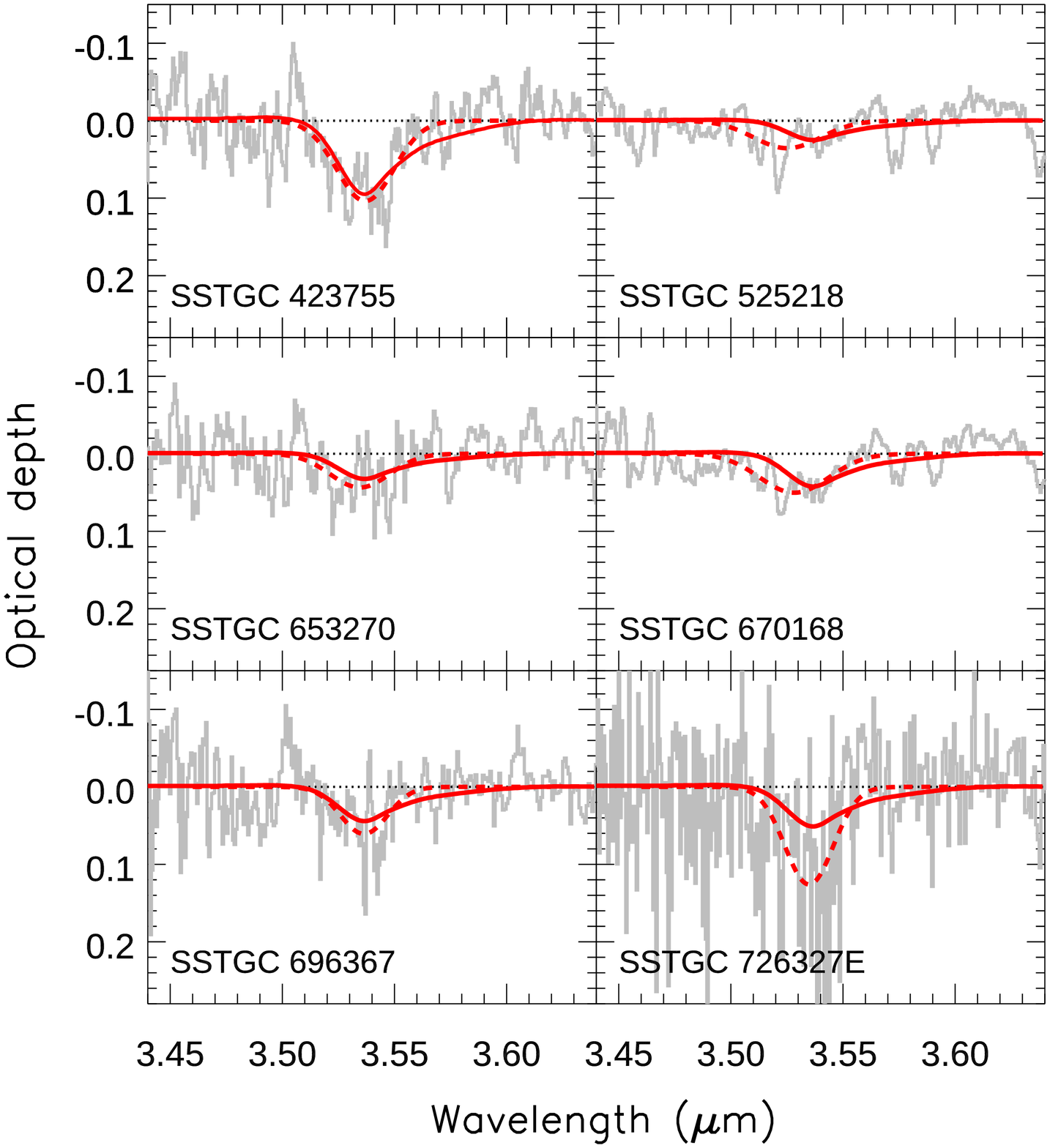}
\caption{Optical depth spectra of sources with mild or weak $3.535\ \micron$ absorption from CH$_3$OH ice. The red line shows the best-fitting model spectrum of pure CH$_3$OH ice grains at $10$~K \citep{hudgins:93}. A independent fit using a gaussian function is shown by a red dotted line.
\label{fig:ch3oh}}
\end{figure}

About half a dozen sources in our sample show mild or weak absorption at $3.535\ \micron$ owing to solid CH$_3$OH. Optical depth spectra of these sources are shown in Figure~\ref{fig:ch3oh}. To compute $\tau_{3.54}$, a peak optical depth at the band center ($3.535\ \micron$), we set a narrow local continuum using a third-order polynomial over $3.40$--$3.50\ \micron$ and $3.60$--$3.70\ \micron$ within the hydrocarbon absorption band complex: the derivation of $\tau_{3.54}$ is independent of the strength and shape of the $3.4\ \micron$ band. Then we fit a laboratory spectrum of pure methanol ices at $10$~K \citep{hudgins:93} to the optical depth spectra in $3.50$--$3.57\ \micron$; see \citet{an:17} for more information on how we extracted the $3.535\ \micron$ component from the model spectrum.

The sixth and eighth columns in Table~\ref{tab:result} contain $\tau_{3.54}$ and a column density of solid CH$_3$OH, respectively. We computed an uncertainty in $\tau_{3.54}$ as a quadrature sum of (i) a fitting uncertainty, (ii) a difference from the case using a straight line for a local continuum (fitted over $3.45$--$3.50\ \micron$ and $3.58$--$3.60\ \micron$), and (iii) a systematic uncertainty in modeling an observed absorption band. The latter represents a difference from the case using a Gaussian function as an alternative approach (red dotted lines in Figure~\ref{fig:ch3oh}).

To be consistent with our previous work \citep{an:17}, we adopted an average integrated band strength $A=(5.95\pm0.65) \times 10^{-18}$~cm~molecule$^{-1}$ for pure methanol ices from \citet{hudgins:93} and \citet{schutte:96}. The uncertainty in $A$ indicates half of the difference between the two studies, and was incorporated in the estimation of the uncertainty in $N {\rm (CH_3OH)}$. Its size is comparable to the change of $A$ in \citet{kerkhof:99} from variations of ice compositions and temperature. Reassuringly, $N {\rm (CH_3OH)}$ for SSTGC~726327E derived in this study ($0.42\pm0.18 \times 10^{18}\ {\rm cm^{-2}}$) is consistent with our previous measurement based on higher resolution spectra, $(0.47\pm0.05) \times 10^{18}\ {\rm cm^{-2}}$ \citep{an:17}.

\section{Results}\label{sec:result}

In this section, we present observed properties of ices and dust in our sample. We begin by inspecting CO absorption at $2.3\ \micron$, which is normally observed in old giant stars (\S~\ref{sec:co}). In a case study for one of the previously identified YSOs in Paper~I (\S~\ref{sec:726327}), we establish a framework for analyzing our spectra, according to which we interpret a $K$-band counterpart to a red mid-IR source as a background giant behind an extended envelope of a YSO or a dense cloud core. In \S~\ref{sec:ice} and \ref{sec:av}, we present the observed distributions of ice and dust grains imprinted on the SpeX spectra, and hunt for new lines of sight with exceptionally large ice column densities that are closely connected to YSOs and/or cloud cores in the CMZ. Finally, we derive chemical compositions of ice mixtures by combining results from this study and Spitzer/IRS observations in Paper~I (\S~\ref{sec:icemix}).

\subsection{Photospheric CO Absorptions}\label{sec:co}

The presence of $2.3\ \micron$ CO band-head absorptions is indicative of relatively cool atmospheres of evolved stars. Some massive YSOs show CO in emission from an inner disk \citep{blum:04,bik:06}, but CO absorptions are rarely observed in YSOs. In this sense, it was unexpected that the Paper~I and \citet{yusefzadeh:09} samples show high fractions of CO absorptions, because these samples have originally been selected as being YSO candidates. Among $12$ targets in the Paper~I sample, four objects show CO absorptions, while three objects are indecisive owing to low SNRs of their $K$-band spectra (Table~\ref{tab:a11}). Likewise, the majority of the photometric YSO candidates in \citet{yusefzadeh:09} show CO in absorption ($24$ objects out of $27$). None of our SpeX targets show CO in emission.

This may indicate a catastrophic failure of a target-selection scheme based on mid-IR photometry to find deeply embedded YSOs. In particular, a ``maybe'' YSO (SSTGC~653270) and a YSO (SSTGC~726327) in Paper~I exhibit CO in absorption. The presence of CO absorption is in conflict with the observed signatures of youth in these objects, such as strong absorption at $15.4\ \micron$ from CO$_2$ ice mixed with CH$_3$OH (postulated at that time, but confirmed later in a subsequent work, including this study) and strong silicate absorptions. More importantly, the detection of $14.97\ \micron$ CO$_2$ $\nu_2 = 1-0$ in SSTGC~726327 indicates that this line of sight passes through warm and dense gas, such as in the circumstellar disk of a protostar. Furthermore, it shows Br$\gamma$ emission, which can originate from strong stellar wind of a massive star or a massive YSO \citep[e.g.,][]{ishii:02}.

The above odd-looking behavior can be naturally explained by a superposition of two sources on the sky, each of which dominates spectral energy distributions in the $K$ and $L$ band, respectively. \citet{morales:17} inspected near-IR sources in high-resolution UKIDSS images and compared them to photometric YSO candidates in the the Spitzer Galactic Legacy Infrared Mid-Plane Survey Extraordinaire (GLIMPSE) catalog \citep{churchwell:09}, discovering that multiple sources are often seen within a $\sim2\arcsec$ Spitzer/IRAC beam. The contamination and/or confusion of a Spitzer mid-IR source would be more problematic toward extremely crowded regions in the CMZ. Because near-IR sources are physically unrelated to a mid-IR source according to this scenario, it implies a profound impact not only on the selection of YSO candidates from a combination of near- and mid-IR photometry, but also on the derivation of physical quantities of YSOs based on the observed spectral energy distribution.

The objects observed in 2017 also exhibit blends of young and old spectroscopic signatures. As they were selected as known or likely red (super-) giants (see \S~\ref{sec:obs}), unlike the other two samples, it was not unexpected to find that all of their spectra show photospheric CO band-head absorption. K or M spectral types for these objects from the template matching reiterate this result (Table~\ref{tab:result}). On the other hand, Br$\gamma$ emission is found in almost half of this sample ($29/70=0.41$): $16$ objects with strong emission, and $13$ additional objects with weak or possible emission features (``maybe'' in the last column of Table~\ref{tab:result}). The presence of such hydrogen recombination lines suggests that these lines of sight partly overlap with gas clouds ionized by hot stars and/or massive YSOs.

\subsubsection{Notes on Sources without CO Absorption}

There are nine targets in our sample that do not show CO band-head absorptions. Six of them (SSTGC~348392, 388790, 405235, 610642, 716531, and 726327W) belong to the Paper~I sample, and three objects (SSTGC~131902, 416141, 719445) are from \citet{yusefzadeh:09}. \citet{geballe:19} also found featureless or nearly featureless $K$-band spectra in the CMZ, which constitute approximately $10\%$ of their sample of reddened point-like objects. They found that most of these objects are likely hot stars or Wolf-Rayet stars deeply embedded in warm dust shells. In agreement with this finding, our template fitting yields higher surface temperatures for these objects with spectral types of $K$. SSTGC~388790 also exhibits Br~$\gamma$ absorption.

\subsection{A Case Study of SSTGC~726327}\label{sec:726327}

SSTGC~726327 was originally selected as a point source from Spitzer/IRAC images ($\sim2\arcsec$ resolution), but then it was resolved into two point-like sources (SSTGC~726327E and 726327W) in the high-resolution $K$- and $L$-band images \citep{an:17}. The bright object (SSTGC~726372E) has been studied in depth in \citet{an:17} based on IRCS spectroscopy at the Subaru telescope. Below we revisit this system in light of our new data on the fainter source (SSTGC~726327W). This observation can be used to help resolve the aforementioned conflict in the young and old spectral signatures.

\begin{figure*}
\centering
\includegraphics[scale=0.8]{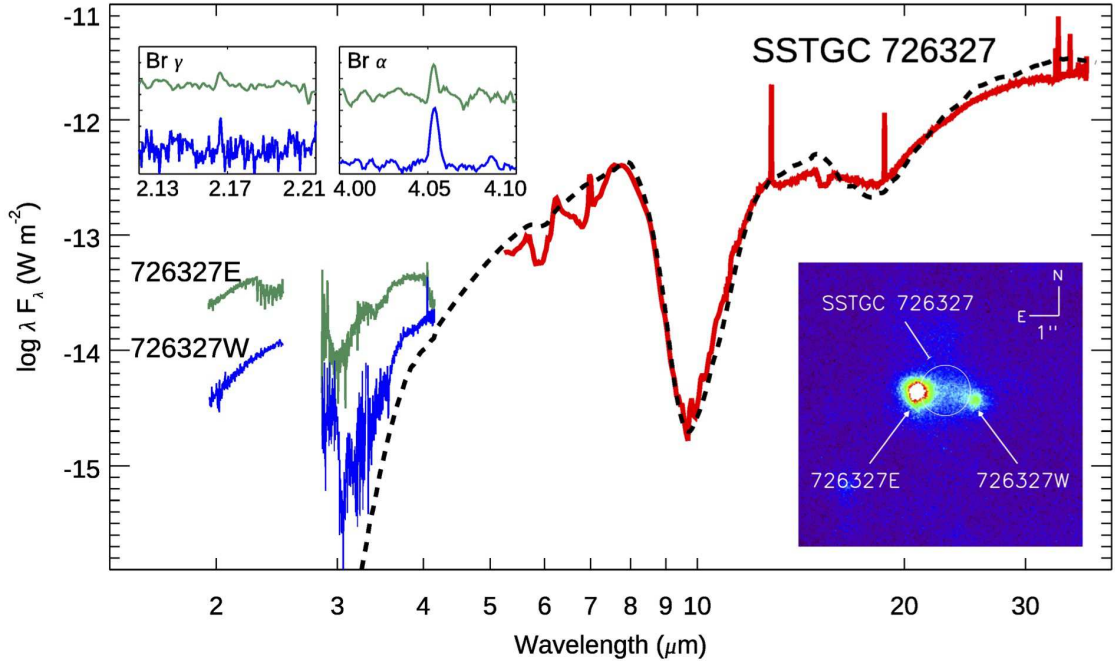}
\caption{Infrared spectra of SSTGC~726327. Originally identified as a single point source in the Spitzer/IRAC image \citep{ramirez:08}, the object was further resolved into two point-like sources (SSTGC~726327E and SSTGC~726327W) in a high-resolution $L'$-band image \citep[][bottom right inset]{an:17}. The IRTF/SpeX spectrum of each source is shown by green and blue solid lines, respectively, while their $5$--$35\ \micron$ Spitzer/IRS spectra \citep{an:11} are displayed by the red line. The black dashed line represents a synthetic spectrum constructed using a composite of $100$~K and $300$~K blackbody curves, a silicate absorption profile from GCS~3 spectrum \citep{kemper:04} and a laboratory spectrum of pure amorphous H$_2$O ice at $T = 10$~K \citep{hudgins:93}. The normalizations, foreground dust extinction ($A_{\rm V}=50$), and H$_2$O ice abundance (corresponding to $\tau_{3.0}=4.0$) are adjusted by hand to provide the most acceptable fit to the IRS spectra.
\label{fig:726327}}
\end{figure*}

Figure~\ref{fig:726327} shows SpeX spectra of each of the two resolved sources, along with $5$--$35\ \micron$ Spitzer/IRS spectra (not resolved into two). Their near-IR spectra are characterized by strong $3\ \micron$ solid H$_2$O and hydrocarbon absorption bands. From our SpeX template fitting, we found $\tau_{3.0}=1.45$ and $A_{\rm V}=33.7$ for SSTGC~726327E. The spectrum of SSTGC~726327W is quite noisy near $3\ \micron$, but the H$_2$O ice band ($\tau_{3.0}=3.5$ and $A_{\rm V}=66.8$) is even stronger than that of SSTGC~726327E. As noted earlier, SSTGC~726327E shows $2.3\ \micron$ CO band-head absorption, but SSTGC~726327W does not. Both objects show Br$\alpha$ and Br$\gamma$ emissions. The Spitzer/IRS mid-IR spectrum of SSTGC~726327 is full of spectroscopic signatures of youth: the boxy CO$_2$ ice band, $14.97\ \micron$ CO$_2$ gas absorption, and ionized gas emissions ($12.8\ \micron$ [\ion{Ne}{2}], $18.7$ and $33.5\ \micron$ [\ion{S}{3}], and $34.8\ \micron$ [\ion{Si}{2}]). Indeed, the source is coincident with a compact \ion{H}{2} region Sgr~B1~A \citep[][GPSR5~0.488-0.028]{mehringer:92,becker:94}. Its IRS spectrum is also characterized by strong silicate bands ($10$ and $18\ \micron$) and water-ice absorption at $12\ \micron$.

As shown by the black dashed line, we constructed a simple synthetic spectrum to describe the overall shape of the observed mid-IR spectrum in the following way. We used two blackbody radiation curves for a continuum and adopted a silicate absorption profile in \citet{kemper:04} and a spectrum from a $10$~K pure H$_2$O ice model in \citet{hudgins:93}. We adjusted parameters (blackbody temperatures and relative strengths of these components) until we achieved a satisfactory eye-ball match to the data. We found that a single blackbody radiation curve is insufficient; instead, a composite of blackbody curves ($\sim100$ and $\sim300$~K) is required to fit the continuum emission at $>5\ \micron$. The necessity of the warm dust component is consistent with our template fitting, in which a blackbody curve with a similar temperature was needed to depict the excess continuum emission at $>4\ \micron$ in the SpeX data. Although our model is not perfect, it shows that the mid-IR continuum can be explained by warm and cold dust emissions.

All of the above evidences and a subsequent detection of methanol ice toward SSTGC~726327E \citep{an:17} suggest that SSTGC~726327E is a background (super-) giant, of which sightline passes through a compact and dense cloud with warm and cold dusts. The spectrum of SSTGC~726327W was too noisy to definitely detect the methanol-ice band. Nonetheless, its stronger H$_2$O ice absorption and larger dust attenuation than SSTGC~726327E suggest that it traces a thicker and dustier envelope of the cloud. In addition, its featureless $K$-band spectrum and strong Br$\alpha$ emission indicate that the source is likely a massive YSO or a hot massive star deeply embedded in its natal molecular cloud. Both objects lie at a projected distance of $\sim8000$--$10,000$~au from the center of the mid-IR source at the distance to the GC ($8$~kpc), which is comparable to the envelope size of a massive YSO \citep[e.g.,][]{vandertak:00}.

Therefore, the simplest explanation for the mixtures of young and old spectroscopic signatures is that the mid-IR source (SSTGC~726327) is likely a {\it foreground} massive YSO, while both near-IR sources (SSTGC~726327E and 726327W) are behind (or inside of) its extended envelope. The above example also leads to a plausible conclusion that our SpeX targets with CO absorptions in their spectra are (super-) giants behind extra dust cloud layers, giving their redder IR colors than numerous field stars in the CMZ. While the source contamination can make a subsequent analysis complicated, it also opens an opportunity to trace ices and dust in foreground intervening clouds using background stars as a distant light source.

\subsection{Ice-rich Sight Lines}\label{sec:ice}

\begin{figure}
\epsscale{2.3}
\plottwo{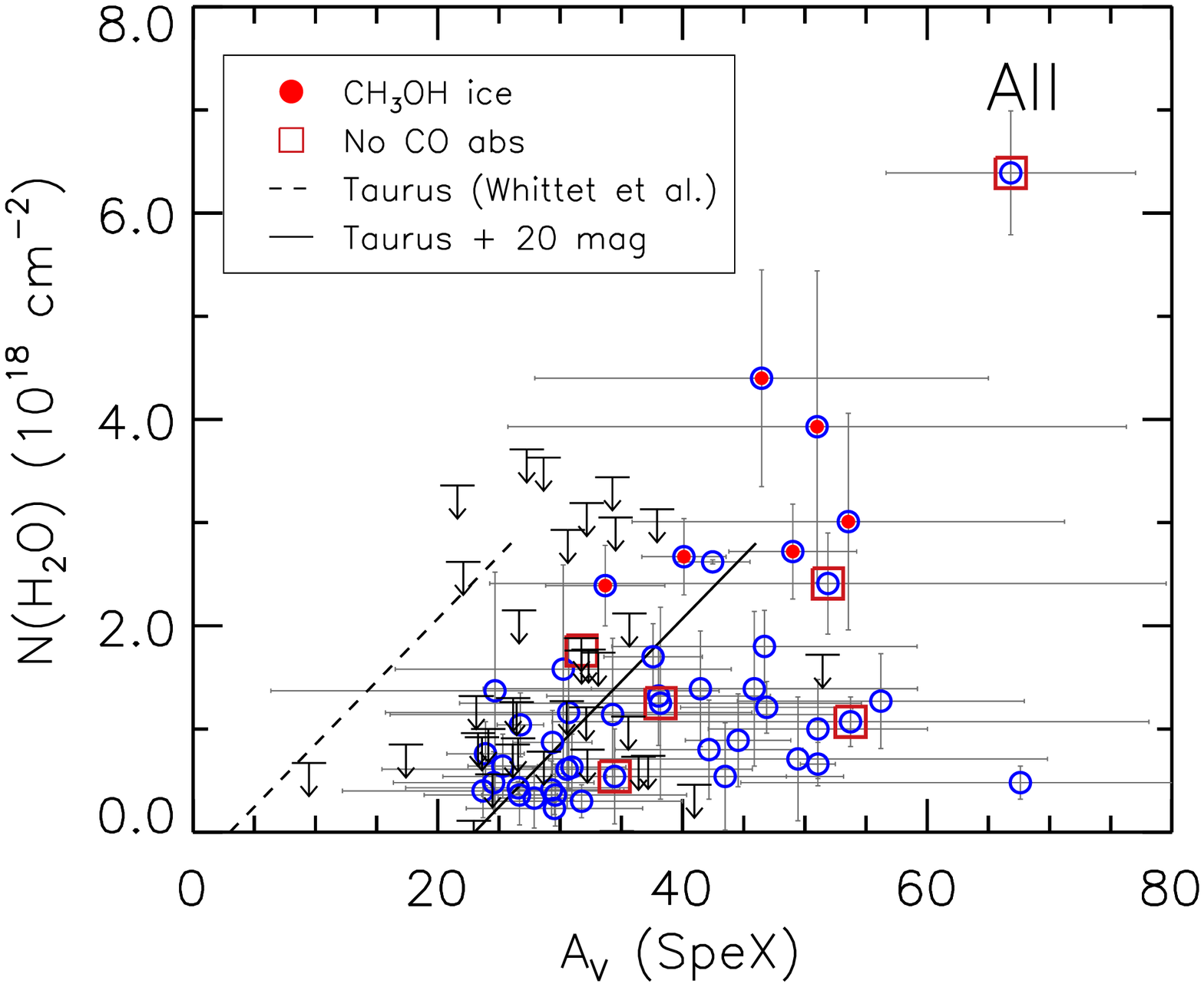}{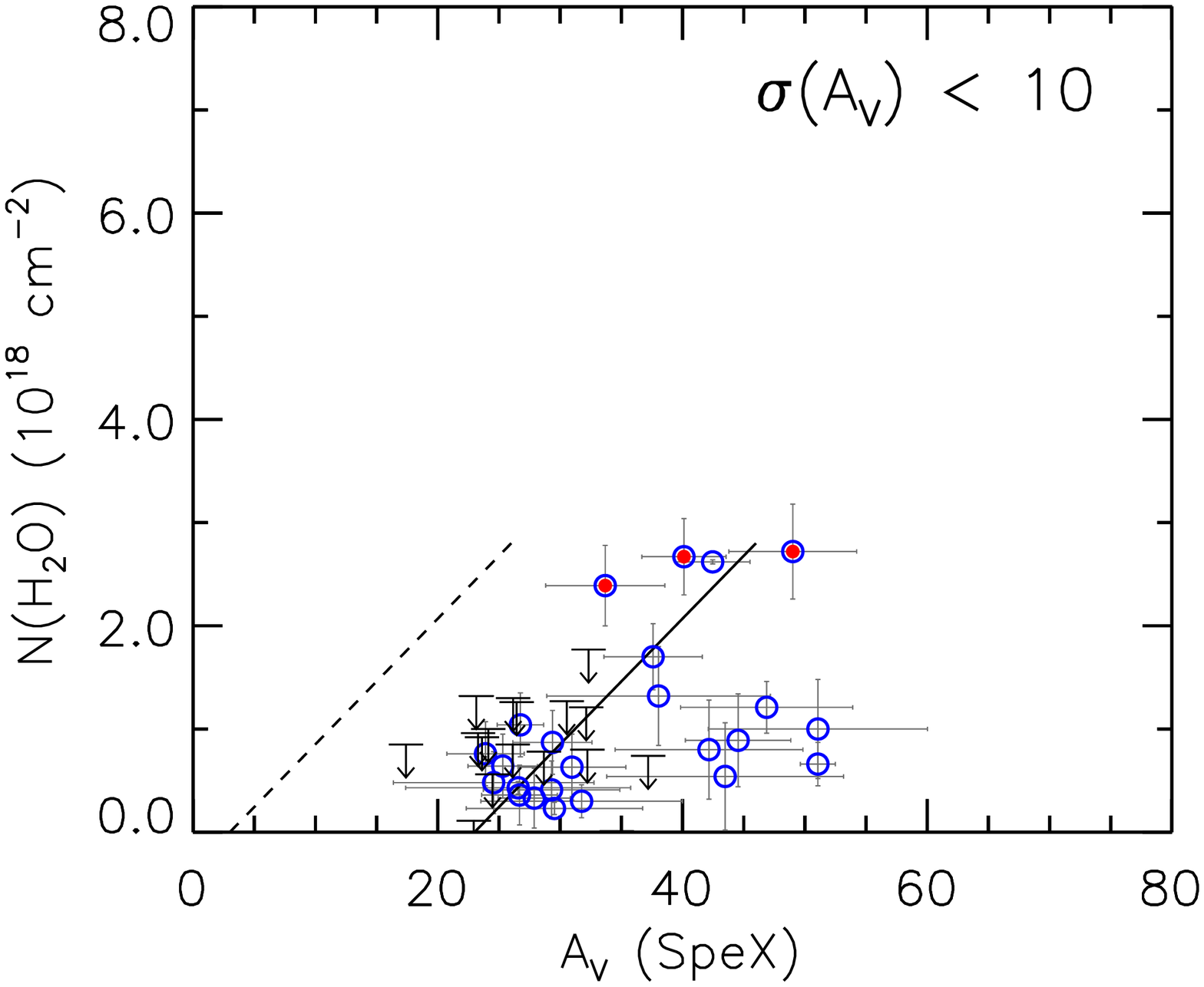}
\caption{The ice column density of H$_2$O against foreground extinction. The dashed lines in both panels indicate a mean relation in the Taurus dark clouds \citep{whittet:01}. The solid lines are the same relation, but shifted horizontally by $\Delta A_{\rm V} = 20$~mag to take into account dust attenuation from foreground diffuse ISM in lines of sight toward the CMZ. The bottom panel is the same as in the top panel, except that only objects with good extinction measurements [$\sigma (A_{\rm V}) < 10$] are included.
\label{fig:h2o}}
\end{figure}

Figure~\ref{fig:h2o} shows $N {\rm (H_2O)}$ against $A_{\rm V}$ for our SpeX targets with $L$-band spectra. For comparison, the dashed line shows a previously known relation in the Taurus cloud \citep{whittet:01}, which has an intercept at $A_{\rm V}=3.2$~mag (corresponding to a formation threshold of $A_{\rm V}=1.6$~mag) and $\Delta \tau_{3.0}/\Delta A_{\rm V} = 0.072$ \citep[see also][]{chiar:11}. The solid line has the same slope as the dashed line, but is shifted by $\Delta A_{\rm V}=20$~mag to take into account foreground dust absorption toward the GC from diffuse ISM \citep{whittet:97}. It was not intended to fit the data, but to understand the observed distribution of the CMZ sources. Albeit with large scatters, the majority of them seem to follow the above shifted line. This is better seen in the bottom panel, where only objects with good $A_{\rm V}$ measurements ($\sigma < 10$~mag) are displayed.

Nevertheless, there are about a dozen sources in the top panel of Figure~\ref{fig:h2o} with systematically larger $A_{\rm V}$ than the solid line by up to $\Delta A_{\rm V} \sim 20$~mag. These objects have intermediate H$_2$O column densities ($\sim 1\times10^{18}$~cm$^{-2}$), indicating that they are probably located behind extra diffuse clouds. Conversely, their presence can also be understood in terms of an elevated threshold extinction, above which H$_2$O ice forms. For example, \citet{eiroa:89} and \citet{tanaka:90} found higher $A_{\rm V}$ in Serpens ($\sim6$~mag) and $\rho$ Ophiuchi ($\ga10$~mag), respectively, of which difference was interpreted as an elevated shielding from stronger UV radiation field in a more active star-forming environment \citep[see discussions in][]{chiar:11}. The same effect may be at work in the CMZ, since the UV radiation field varies across the region \citep{an:13,simpson:18}.

While our sample covers a wide range of the H$_2$O ice absorption, we particularly note that some of the $L$-band spectra are carved by its exceptionally strong and wide absorption. In the top panel of Figure~\ref{fig:h2o}, there are nine sources with the strongest H$_2$O ice absorptions with $N {\rm (H_2O)} > 2\times10^{18}\ {\rm cm}^{-2}$ ($\tau_{3.0} > 1.2$) that are clearly distinct from the rest of the objects. Five of them are from Paper~I (SSTGC~610642, 653270, 696367, 726327E, and 726327W), and two are from the observing run in 2017 (SSTGC~423755 and 670168). The rest are GCIRS~7 (SSTGC~525210) and GCIRS~19 (SSTGC~528212). Their Spex spectra are shown in Figure~\ref{fig:spex}. The low fraction of strong H$_2$O ice sources in the 2017 sample implies that the lines of sight with strong ice-rich absorptions are not common in the CMZ.

\begin{figure}
\epsscale{2.3}
\plottwo{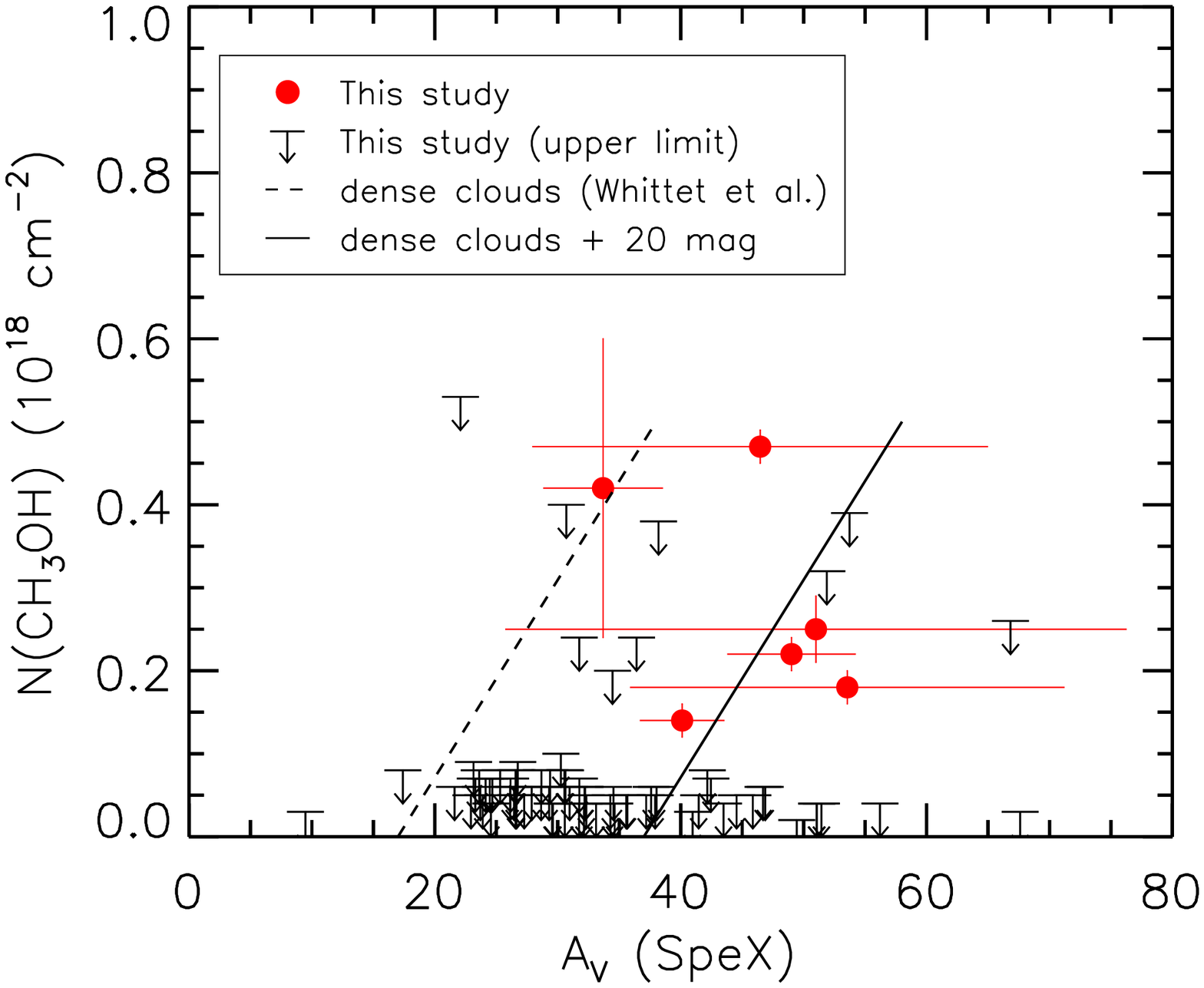}{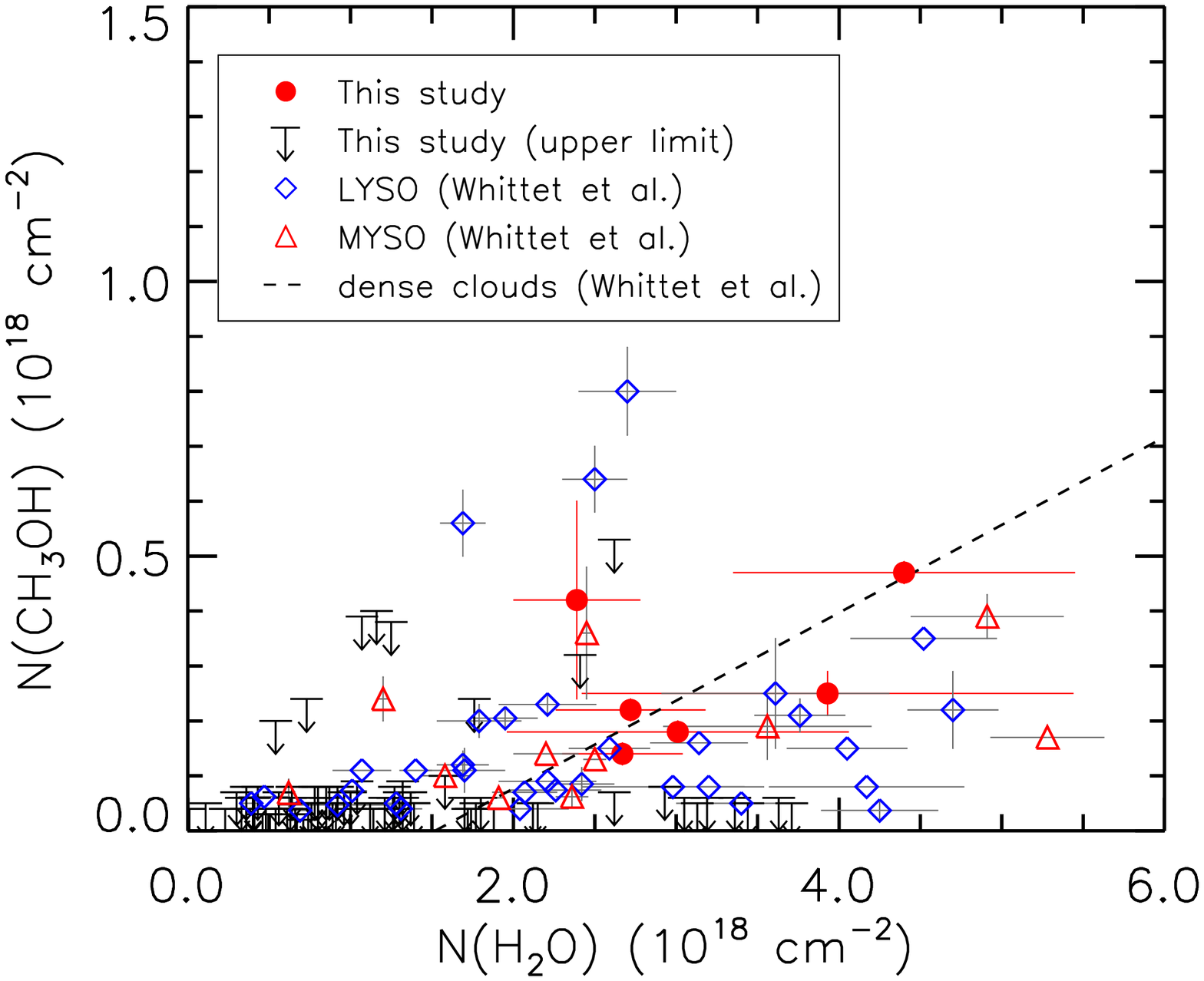}
\caption{Top: the ice column density of CH$_3$OH against foreground extinction. The dashed line indicates a mean relation from field stars behind dense molecular clouds \citep{whittet:11}, while the solid line represents the same slope, but at a higher threshold extinction with $\Delta A_{\rm V} = 20$ to take into account the foreground extinction in the Galactic disk toward the CMZ. Bottom: the CH$_3$OH ice against H$_2$O ice column densities. Open symbols represent measurements toward low-mass (LYSO; blue diamonds) and massive YSOs (MYSO; red triangle) in \citet{whittet:11}. The dashed line indicates their mean relation from field stars behind quiescent, dense molecular clouds.
\label{fig:ch3oh2}}
\end{figure}

In addition to H$_2$O ice, we detected methanol-ice absorption from six targets in our sample (\S~\ref{sec:ch3oh}), as shown by red filled circles superimposed on open circles in Figure~\ref{fig:h2o}. Three of them are from the Paper~I sample, two are the sources observed in 2017, and the remaining is GCIRS~19. All of them show large foreground dust attenuation ($A_{\rm V} > 30$) and strong H$_2$O ice absorption ($N > 2\times10^{18}\ {\rm cm}^{-2}$). Their CH$_3$OH ice column densities against foreground extinction are shown in the top panel of Figure \ref{fig:ch3oh2}. As in the case of H$_2$O ice, methanol ices are generally observed toward regions with high foreground extinction, but with a significantly elevated $A_{\rm V}$ threshold. The dashed line represents a previously studied relation from field stars behind various dense molecular clouds \citep{whittet:11}. The solid line has the same slope, but with a horizontal shift by $\Delta A_{\rm V} = 20$ to take into account the extinction from foreground diffuse ISM toward the CMZ \citep{whittet:97}. Again, this was not intended to fit the data, but a fair agreement of the SpeX data with the solid line indicates that the formation threshold and the growth of CH$_3$OH ice grains in the CMZ are similar to those in the local environment.

In the bottom panel of Figure~\ref{fig:ch3oh2}, the abundance distribution of solid CH$_3$OH with respect to solid H$_2$O is shown by red filled circles. The open symbols are measurements toward low-mass YSOs (blue diamonds) and massive YSOs (red triangles) in \citet{whittet:01}, while the dashed line is their average relation observed toward field stars behind various dense molecular clouds. Several upper limits in their measurements are not included in Figure~\ref{fig:ch3oh2}. Their data indicate that there is a significant scatter in $N {\rm (CH_3OH)} / N {\rm (H_2O)}$ with no clear division between different classes of the objects \citep[see also][]{boogert:15}. Nonetheless, we note that our IRTF objects fall into a region, which is occupied by the sight lines toward dense molecular clouds and YSO envelopes in the above study.

All of the six objects with CH$_3$OH ice detection have the strongest H$_2$O ice absorption ($N > 2\times10^{18}\ {\rm cm}^{-2}$ or $\tau_{3.0} > 1.2$). Only upper limits in $N {\rm (CH_3OH)}$ are available for the remaining three objects [SSTGC~525210 (GCIRS~7), 610642, and 726327W]. CH$_3$OH ice traces YSO envelopes and dense media, all of which may subsequently turn into a collapse of a prestellar core. Therefore, the above condition ($\tau_{3.0} > 1.2$) seems intimately related to a star-formation activity in the CMZ. In support of this argument, \citet{tanaka:90} studied the $\rho$ Ophiuchi star-forming region, and found a similar gap in the $\tau_{3.0}$ distribution, which distinguishes protostars ($\tau_{3.0} \ga 1$) from early-type and emission-line stars in the cloud ($\tau_{3.0} \la 0.3$).

\subsection{Excess Amount of Foreground Dust}\label{sec:av}

\begin{figure}
\epsscale{2.3}
\includegraphics[scale=0.36]{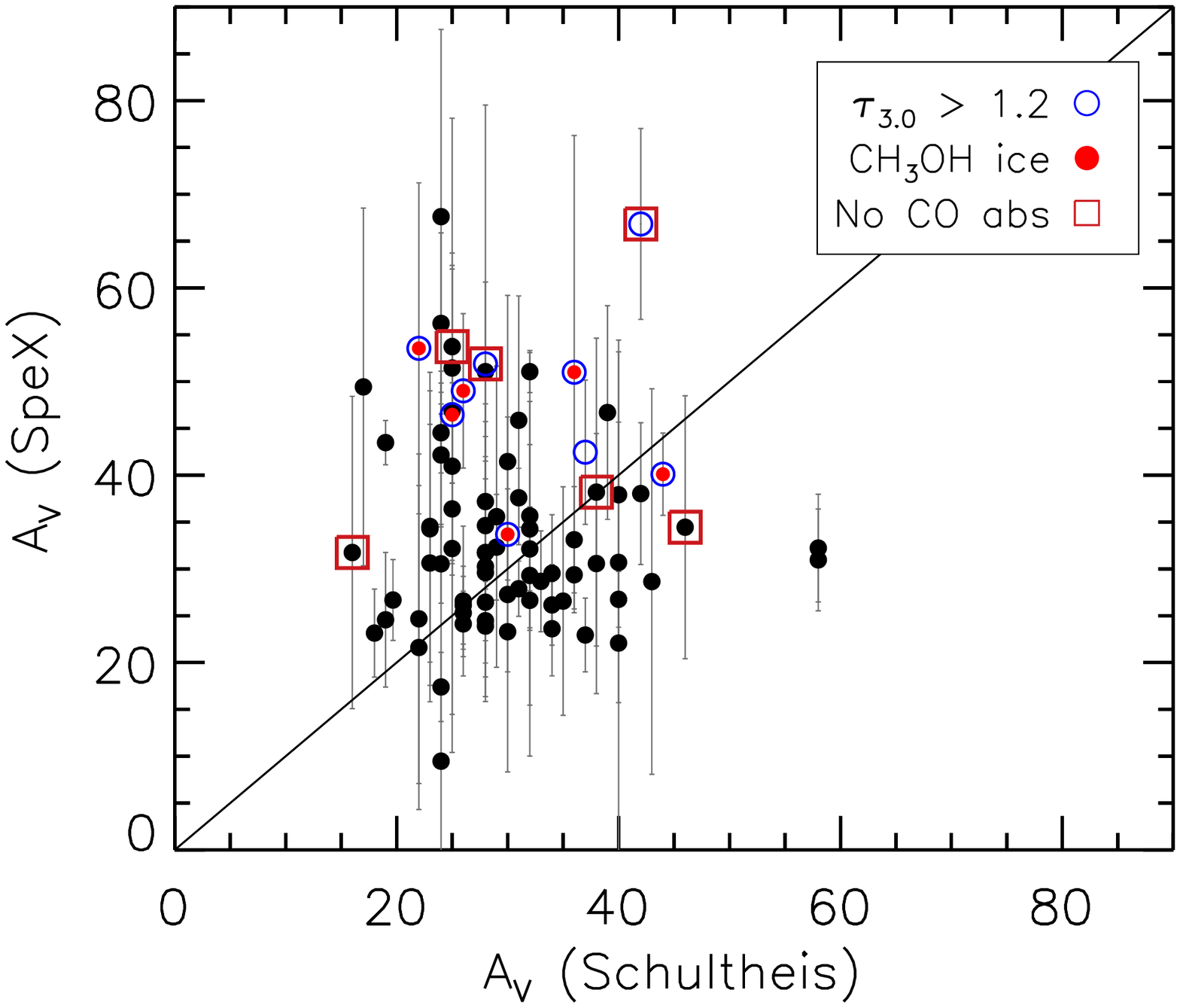}
\includegraphics[scale=0.36]{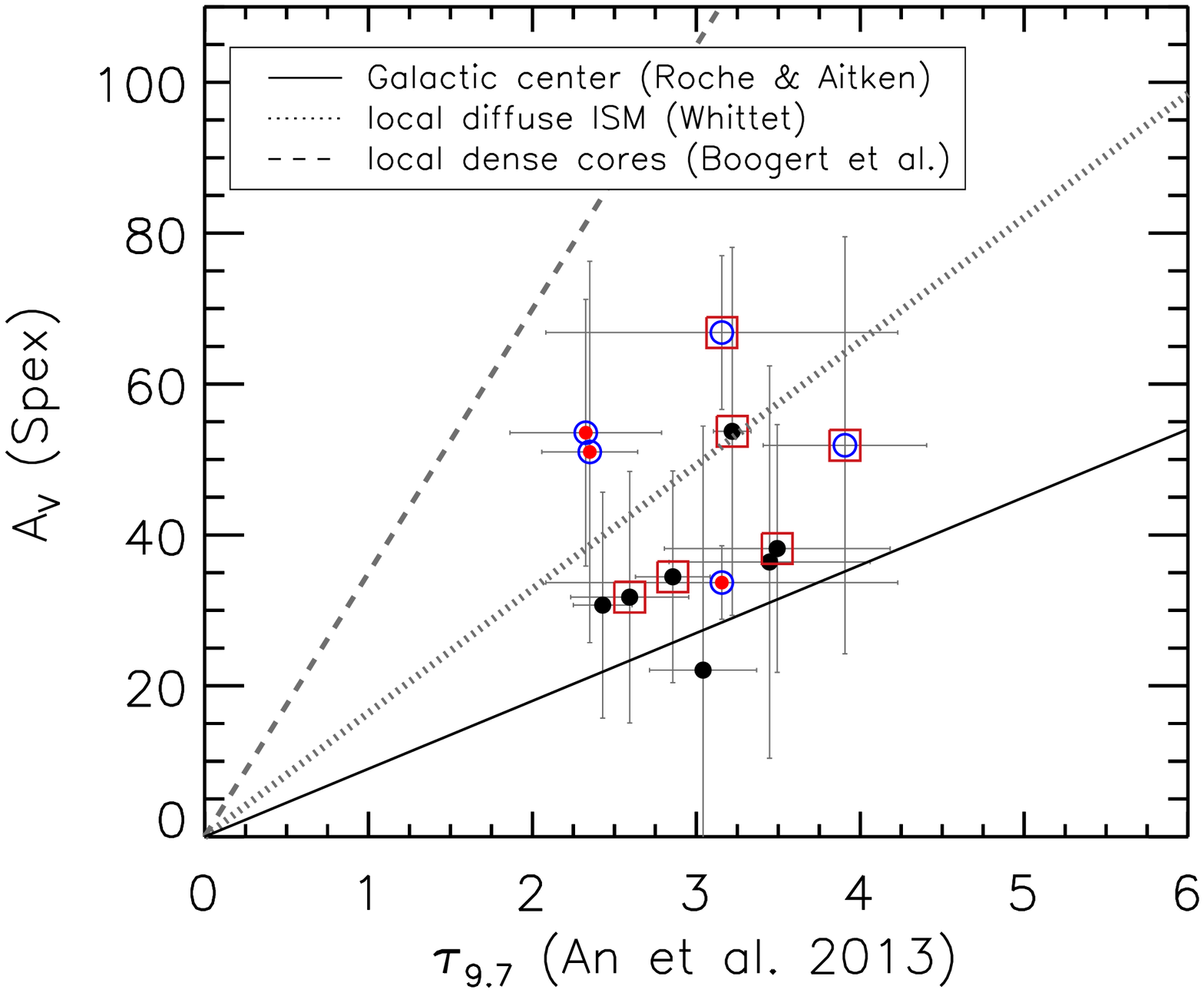}
\includegraphics[scale=0.36]{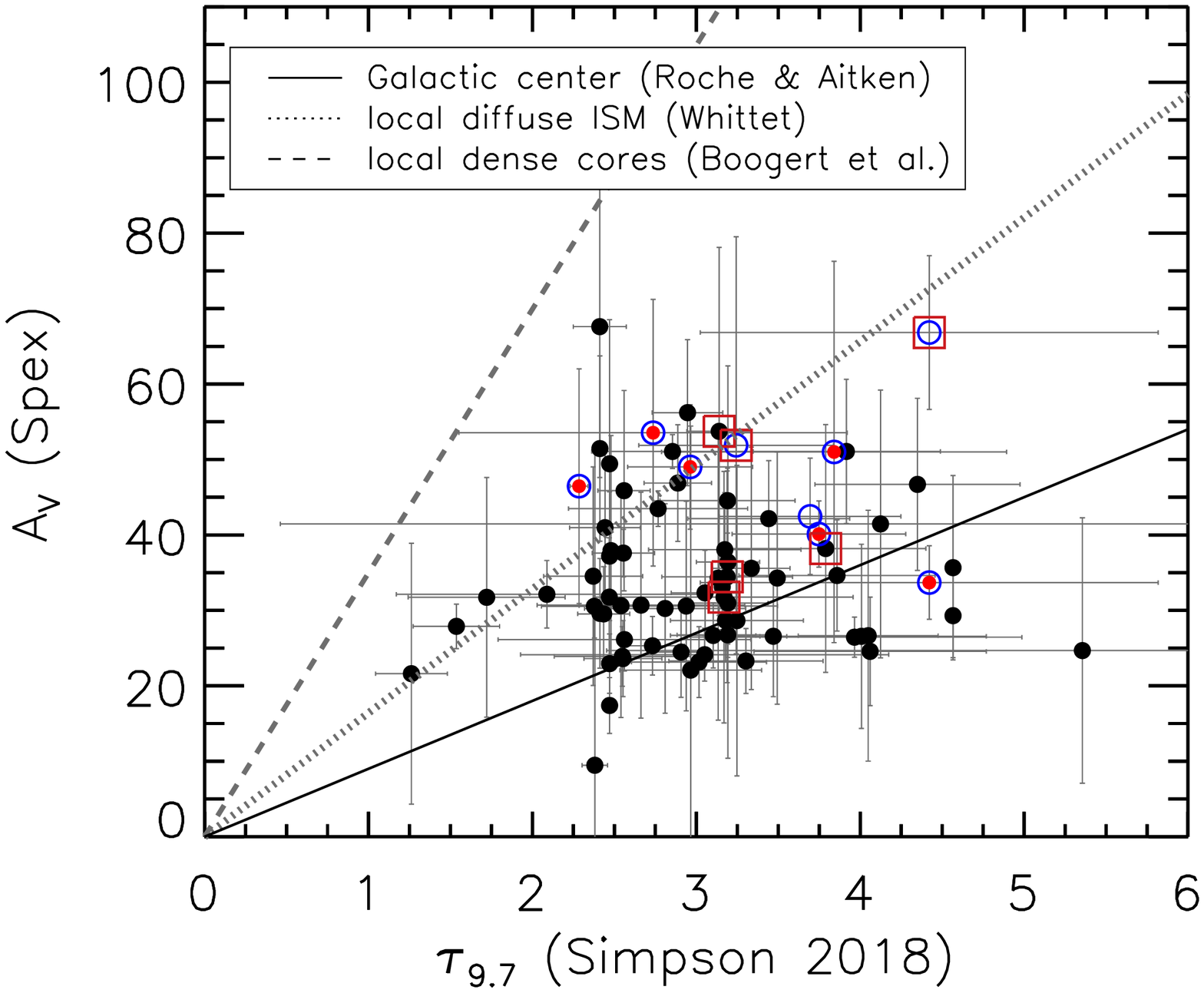}
\caption{Top: comparison of foreground extinction estimates based on SpeX spectra (this study) with extinction-map values from colors of red giant stars \citep{schultheis:09}. A unity is indicated by a solid line. Middle: comparison with extinction estimates in \citet{an:13} based on the depth of $10\ \micron$ silicate absorption band ($\tau_{9.7}$) in adjacent clouds. Only those based on high-resolution IRS spectra are shown. Bottom: same as in the middle panel, but based on $\tau_{9.7}$ measurements in \citet{simpson:18}.
\label{fig:av}}
\end{figure}

Figure~\ref{fig:av} shows a comparison of foreground extinction estimates from SpeX spectra with external measurements in the literature. In the top panel, SpeX-based estimates are compared to those from the extinction map in \citet{schultheis:09}, which is based on mean 2MASS and IRAC colors of red giant stars in a grid of $2\arcmin \times 2\arcmin$. The scatter in Figure~\ref{fig:av} is large ($\sigma = 10$~mag), which is in part due to large uncertainties in our measurements. However, it could also be caused by variable foreground extinction in the CMZ over a length scale smaller than the grid size of the extinction map.

In the top panel of Figure~\ref{fig:av}, objects with a strong H$_2$O ice band ($N > 2\times10^{18}\ {\rm cm}^{-2}$) are shown by blue open circles ($N=9$). Six of them also show CH$_3$OH ice absorption as marked by red filled circles. Our SpeX-based extinction toward these ice-rich objects is systematically higher than the measurements in \citet{schultheis:09}. Their mean extinction is $\langle A_{\rm V} \rangle = 48\pm3$~mag, which is significantly higher than the average extinction toward all of our sample in \citet{schultheis:09}, $\langle A_{\rm V} \rangle = 30.3$~mag. On the other hand, the mean extinction from our template matching is more consistent with this average value, if the ice-rich sources are removed ($\langle A_{\rm V} \rangle = 35\pm1$~mag). This indicates that objects with large $N {\rm (H_2O)}$ lie behind extra dust layers, unlike numerous field stars in the CMZ.

The middle panel of Figure~\ref{fig:av} displays a comparison between SpeX-based $A_{\rm V}$ and $\tau_{9.7}$ (a peak optical depth of the $9.7\ \micron$ silicate feature) in \citet{an:13}. We only include Paper~I sources in this panel, because their ``background'' ISM spectra are readily available from our Spitzer/IRS observations, based on which $\tau_{9.7}$ were computed from a flux ratio between $10\ \micron$ and $14\ \micron$ of photodissociation regions \citep{simpson:07}. We took an average of $\tau_{9.7}$ measurements from a set of the ISM spectra around each source with $\sim\pm1\arcmin$ offsets (see Paper~I for more details) and therefore $\tau_{9.7}$ provides ``average'' foreground extinction in a $\sim2\arcmin\times2\arcmin$ field in front of photodissociation regions (like in the case of red giants in the top panel, but in an independent way). The horizontal error bars show a standard deviation of individual $\tau_{9.7}$ from the background ISM spectra.

Likewise, the bottom panel shows a comparison with $\tau_{9.7}$ measurements in \citet{simpson:18}, which is essentially based on the same technique as in \citet{an:13}. The extinction map in \citet{simpson:18} is also largely based on our Spitzer/IRS data in Paper~I, and therefore the two studies are not entirely independent from each other. On the other hand, more objects are found in the bottom panel, because \citet{simpson:18} employed low-resolution, long-slit IRS spectra to cover much larger areas in the CMZ (as opposed to those based on high-resolution data in the middle panel). We estimated $1\sigma$ uncertainties in $\tau_{9.7}$ by computing a standard deviation of measurements in a $\pm1\arcmin\times\pm1\arcmin$ grid centered on each object. There are a few sources without error bars, owing to a coarse grid of the extinction map in such regions.

In the middle and bottom panels of Figure~\ref{fig:av}, the solid line indicates the previously established $A_{\rm V}$ versus $\tau_{9.7}$ relation toward the GC \citep{roche:85}, while the dotted line represents the relation from diffuse ISM \citep{whittet:03}, which is about a factor of two steeper than the GC relation \citep[see also][]{chiar:06}. In the bottom panel, most of the objects fall into a region, loosely confined by these two relations. Although it is not straightforward to assign individual objects to each group, it is interesting to note that most ice-rich sight lines (blue open circles or red dots) are placed above the GC line. The same bias can also be seen in the middle panel. Dense cloud cores in the Galactic disk exhibit weaker silicate absorptions for a given visual or near-IR extinction \citep[][dashed line]{boogert:13}, probably due to enhanced grain growth in dense media \citep[see also][]{chiar:07}. If the same mechanism is at work in the CMZ, the systematically larger $A_{\rm V}$ of the ice-rich sight lines for a given amount of silicates than the mean GC line (conversely, smaller $\tau_{9.7}$ for a given $A_{\rm V}$) implies extra dense clouds in front of these objects.

\begin{figure}
\epsscale{2.3}
\plottwo{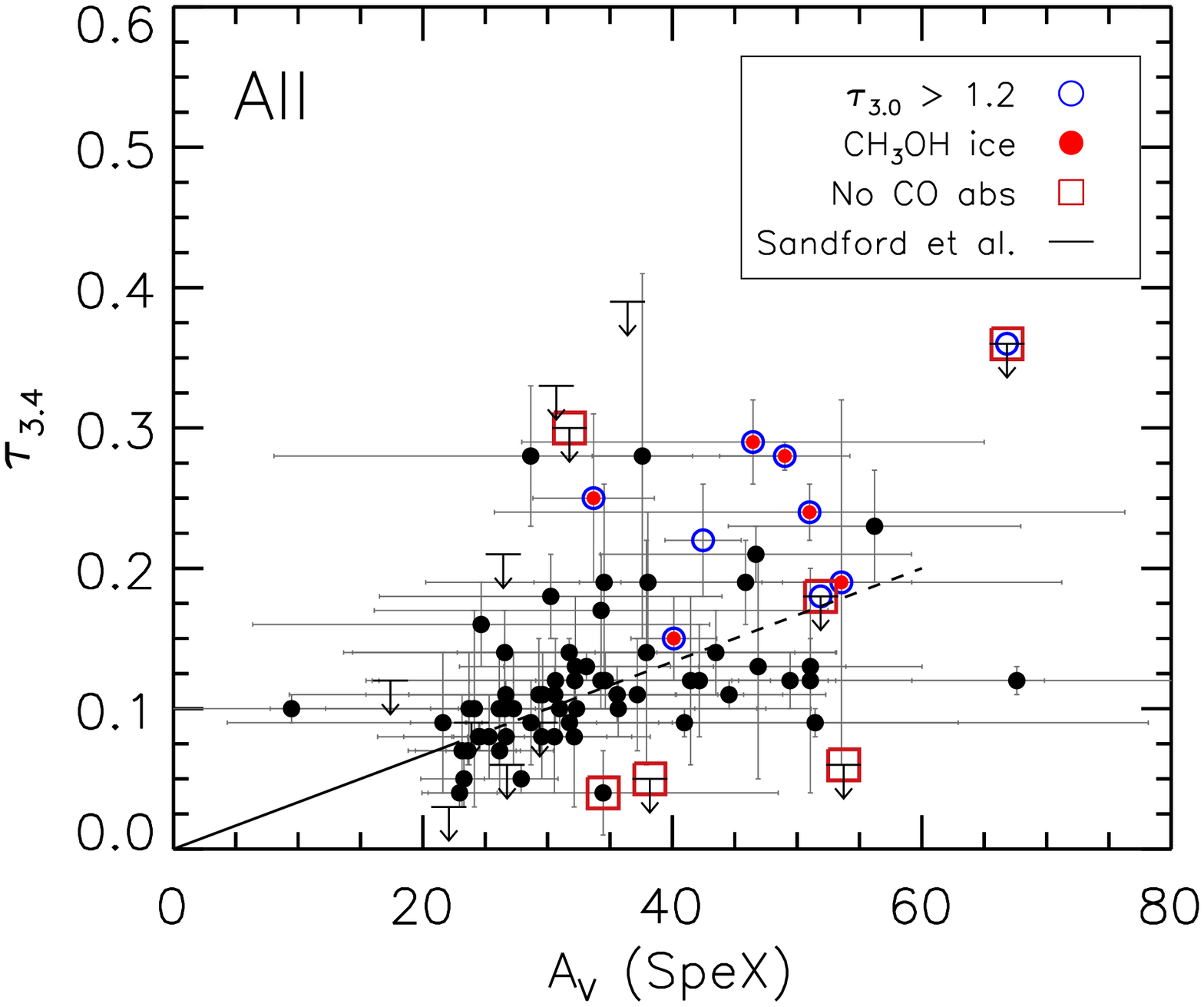}{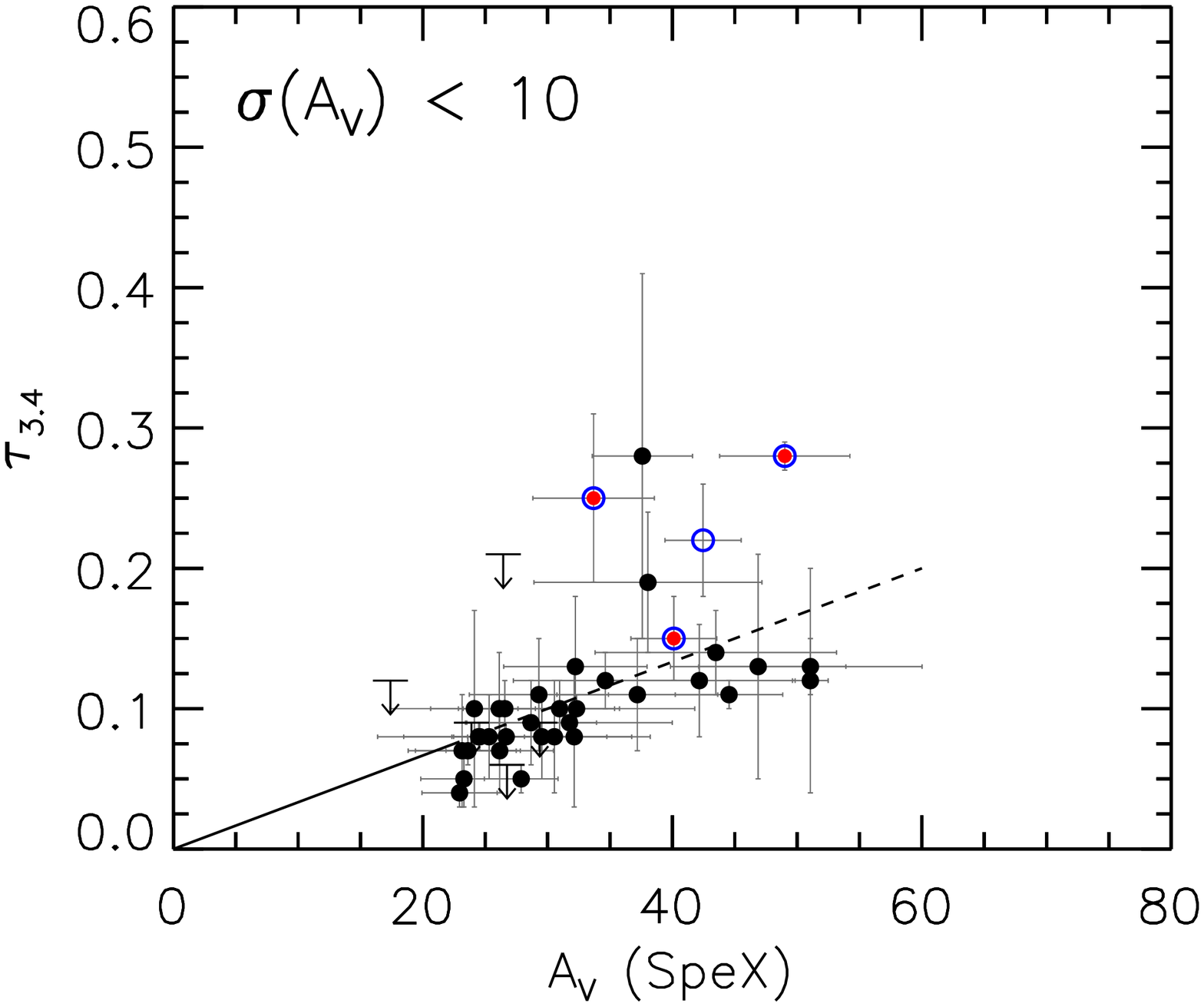}
\caption{Foreground extinction versus the peak optical depth at $3.4\ \micron$ from hydrocarbon absorption. The solid line represents the relation obtained from the local, diffuse ISM \citep{sandford:95}, while its projection beyond the upper limit in their sample ($A_{\rm V} > 22$) is shown by a dashed line. The bottom panel is the same as in the top panel, except that only those with good extinction measurements are shown.
\label{fig:av_34}}
\end{figure}

In addition to the above foreground extinction estimates, the broad absorption at $3.4\ \micron$ from C--H stretch of aliphatic hydrocarbons were often employed as a tracer of interstellar dust grains in diffuse ISM \citep{sandford:91,sandford:95,pendleton:94}. Figure~\ref{fig:av_34} shows the peak optical depth of the $3.4\ \micron$ band ($\tau_{3.4}$) against $A_{\rm V}$ as derived from our spectral matching. In local diffuse clouds, a tight correlation exists between $\tau_{3.4}$ and a foreground extinction, as shown by the solid line at $A_{\rm V} < 25$ \citep[$\langle A_{\rm V}/\tau_{3.4} \rangle \approx 240$;][]{sandford:95}. Because our sample was selected as those likely behind a large amount of clouds, there are only few objects at $A_{\rm V} < 25$. Nonetheless, Figure~\ref{fig:av_34} shows, based on an extensive data set in this study, that ``normal'' GC sources with weak or intermediate ice absorptions (black dots) follow its extrapolated relation (dashed line). This trend can be better seen in the bottom panel, where an observational scatter is reduced by including only those with small $A_{\rm V}$ measurement uncertainties.

In the top panel of Figure~\ref{fig:av_34}, about half of the objects with strong $3.4\ \micron$ absorptions ($\tau_{3.4} > 0.15$) have significant ice absorptions (blue open circles and red filled circles), while some of ``normal'' GC sources are also found at large $\tau_{3.4}$. The large overlap between sources with and without water ices agrees with our earlier conclusion drawn from Figure~\ref{fig:av} that a large amount of dust grains does not necessarily indicate a large ice column density in the extremely patchy CMZ region. Nevertheless, ice-rich objects show a skewed $\tau_{3.4}$ distribution toward higher values with respect to the dashed line, whereas objects with weak or intermediate ice absorptions are more evenly distributed.

\begin{figure}
\epsscale{1.2}
\plotone{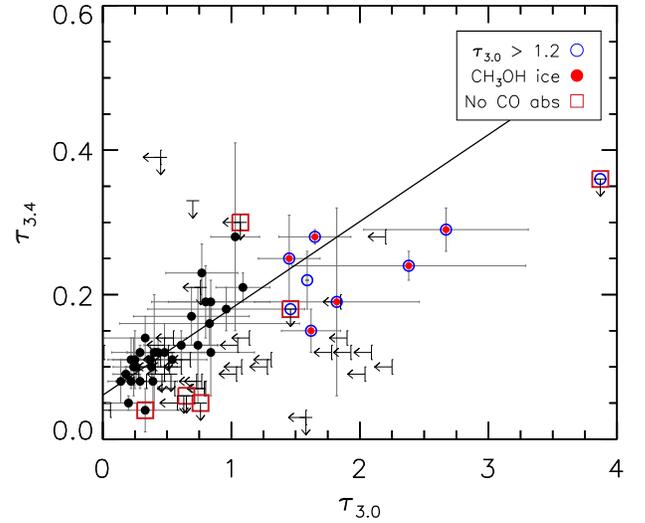}
\caption{The peak optical depth at $3.4\ \micron$ versus the peak optical depth at $3.0\ \micron$. The solid line represents a mean relation from a linear least-squares fit with weights given by error bars on both axes.
\label{fig:h2o_hc}}
\end{figure}

Figure~\ref{fig:h2o_hc} displays $\tau_{3.4}$ against $\tau_{3.0}$. Although the scatter of observational data is large, systematically stronger hydrocarbon absorptions toward sources with larger ice column densities can be seen. The solid line represents a linear least-squares fit to the data, which took into account uncertainties in both axes: $\tau_{3.4} = (0.12\pm0.01) \tau_{3.0} + (0.06\pm0.01)$. Previously, \citet{moultaka:04,moultaka:15} reported a similar positive correlation in the central $0.5$~pc of the GC (with a slope of $0.03$--$0.17$), arguing that the ISM adjacent to Sgr~A$^*$ is a mixture of dense and diffuse media. Our SpeX data cover a significantly larger volume than in \citet{moultaka:04,moultaka:15}, and also include a larger number of objects with significantly enhanced water-ice absorption. The positive correlation between $\tau_{3.0}$ and $\tau_{3.4}$ found in the above studies and this work point to the fact that carriers of the $3.4\ \micron$ absorption are not only present in a diffuse ISM, but also in part in a dense medium, unlike previous claims that they only trace diffuse ISM \citep[e.g.,][]{mcfadzean:89,sandford:91}.

The above result rephrases our conclusion based on Figure~\ref{fig:av} that ice-rich sight lines are generally more obscured by foreground dust than typical CMZ sources. Nonetheless, there is a tendency for the average $\tau_{3.4}$ to level off at large ice column densities ($\tau_{3.0} > 1.2$). The saturation of the $\tau_{3.4}$ versus $\tau_{3.0}$ relation may indicate that the formation of $3.4\ \micron$ carriers are suppressed somehow in dense molecular clouds. For example, \citet{greenberg:95} found in a laboratory experiment that aliphatic hydrocarbons can form by strong ultraviolet (UV) irradiation on ice mantles. Since dense clouds are shielded from UV photons, a dearth of hydrocarbon grains in YSO envelopes or dense molecular cores is expected, leading to the saturation of a hydrocarbon column density. It may also be that, as ice grains grow in dense media, more carbonaceous dust grains are coated with ice mantles. However, the current data are insufficient to discriminate between the two scenarios and to judge whether carbonaceous dust grains are more intimately mixed with ice-coated grains throughout the entire volume of the CMZ.

\subsection{Ice Compositions}\label{sec:icemix}

\begin{figure*}
\epsscale{1.1}
\plotone{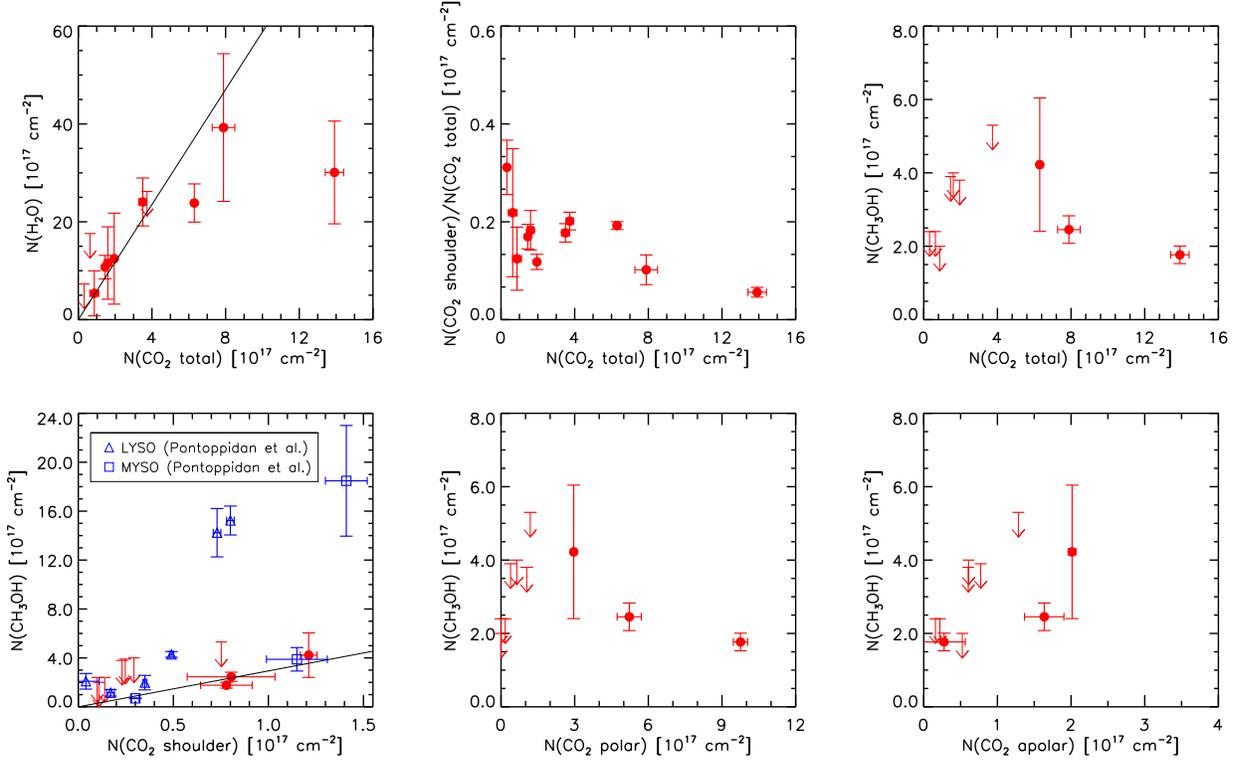}
\caption{Relative abundances of H$_2$O--CO$_2$--CH$_3$OH ice mixtures in the Paper~I sample. Column densities of solid water [$N$(H$_2$O)] and solid methanol [$N$(CH$_3$OH)] are from this work, while column densities of CO$_2$ ice [$N$(CO$_2$)] are taken from Paper~I based on the $15\ \micron$ absorption band. Among sources in the Paper~I sample, only three objects (SSTGC~653270, 696367, 726327E) have $N$(CH$_3$OH). Top panels: $N$(H$_2$O),  a fraction of the shoulder CO$_2$ ice, and $N$(CH$_3$OH) with respect to the total column density of CO$_2$ ice (a sum of all individual components including the pure and diluted components). Bottom panels: $N$(CH$_3$OH) as a function of column densities of individual CO$_2$ components (the shoulder, polar, and apolar, respectively). The solid line in the top left panel indicates a mean relation in \citet{gerakines:99}. In the bottom left panel, the solid line shows a least-squares fit to our measurements (red filled circles).
\label{fig:ch3oh3}}
\end{figure*}

Figure~\ref{fig:ch3oh3} compares methanol-ice column densities derived in this study to those of solid CO$_2$ from the mid-IR Spitzer/IRS observations in Paper~I. The top left panel shows $N {\rm (H_2O)}$ versus $N {\rm (CO_2)}$ for all of the Paper~I sources included in this study. The $N {\rm (H_2O)}$ was measured from our $3\ \micron$ spectroscopy, while $N {\rm (CO_2)}$ represent those of the total CO$_2$ ice absorption at $15\ \micron$. The solid line indicates a mean relation in \citet{gerakines:99}, who found a tight correlation ($0.17\pm0.03$) in quiescent clouds and massive YSOs, irrespective of their environmental differences. As expected from the tightness of this relation, most of our data are found on top of this relation. The only exception is the one with the strongest CO$_2$ absorption (SSTGC~653270), but it may be partly due to a difference in the aperture size between IRTF/SpeX and Spitzer/IRS. The fraction of the shoulder CO$_2$ ice component with respect to $N {\rm (CO_2)}$ is shown in the top middle panel.

The rest of panels in Figure~\ref{fig:ch3oh3} show $N {\rm (CH_3OH)}$ against total and individual components of solid CO$_2$ in Paper~I: total (top right), shoulder (bottom left), polar (bottom middle), and apolar (bottom right) ices. Only three objects in the Paper~I sample have a clear detection of $3.535\ \micron$ absorption (Fig.~\ref{fig:ch3oh}). Even with this small number of samples, it is readily seen that the strong CO$_2$ shoulder absorption is observed toward sources having a large amount of solid CH$_3$OH. A provisional linear fit to this data suggests that the relative abundance of CH$_3$OH with respect to CO$_2$ is $N$(CH$_3$OH)/$N$(shoulder CO$_2$) $=2.9\pm0.6$. Our best-fitting slope corresponds to a lower bound of earlier estimates in \citet{pontoppidan:08}, who found a wide range of $N$(CH$_3$OH)/$N$(shoulder CO$_2$) between $\sim3$ and $20$ in low- and high-mass YSOs (blue open symbols).

On the other hand, the observed correlations between $N {\rm (CH_3OH)}$ and the other two CO$_2$ components are more subtle. There is a tendency of a decreasing amount of H$_2$O-rich polar ice with $N {\rm (CH_3OH)}$ (bottom middle), while the opposite trend is observed for CO-rich apolar ices (bottom right). Although there are only three data points in this abundance estimation, the apparent trend in the bottom right panel can be understood on the basis of CH$_3$OH formation in ice grain surfaces \citep{watanabe:03,cuppen:09,coutens:17}: the positive correlation implies that CO molecules subsequently react with hydrogen atoms to form CH$_3$OH molecules, as more CO become available in grain surfaces. Furthermore, a non-zero intercept of the trend implies that CH$_3$OH molecules are already in place in abundance even in the absence of CO-rich CO$_2$ ice mantles. This suggests that the formation channel of CH$_3$OH is limited by available CO in grain surfaces, while atomic hydrogens are constantly supplied from dissociation of H$_2$ in molecular clouds by cosmic rays \citep[e.g.,][]{whittet:11}.

\section{Summary and Discussions}\label{sec:summary}

\subsection{Seeing Through the Envelope}

To study the demographics and characteristics of interstellar ices in the CMZ, we conducted $K$- and $L$-band spectroscopic observations of $109$ red point-like sources toward the CMZ using IRTF/SpeX. Our spectroscopic targets were selected as having heavy foreground dust attenuation, based on their distinctively red IR colors. In the absence of direct distance measurements, their high foreground extinctions were taken as a safeguard to put them at the GC distance.

The majority of our SpeX spectra show CO band-head absorption, while they exhibit varying amounts of $3\ \micron$ absorption from solid H$_2$O. We interpret this as evidence that they are background (super-) giants behind molecular clouds in different lines of sight toward the CMZ. In particular, among $82$ sources observed in both $K$ and $L$ band, we detected $3.535\ \micron$ absorption from solid CH$_3$OH in six objects. These sources also show significant concentrations of H$_2$O ice compared to the rest of the objects in our sample. Since CH$_3$OH ices are generally seen toward the densest part of molecular clouds, the low fraction of such objects ($6/82$) strongly argues against a widespread distribution of intervening dense molecular clouds in the CMZ. Instead, we attribute the extra absorption to compact ice-rich clouds, such as an extended envelope of a YSO or a dense molecular cloud core, where star formation has just occurred or is imminent. In support of this conclusion, we found that these ice-rich objects exhibit systematically larger foreground extinction than their neighboring sources on the sky.

\subsection{Composition of Methanol-rich Ices in the CMZ}

Among sources with methanol ice, SSTGC~653270, 696367, and 726327E have mid-IR spectra from Paper~I, from which we extracted a column density of the ``shoulder'' CO$_2$ ice (\S~\ref{sec:icemix}). A proposition that the $15.4\ \micron$ shoulder absorption from solid CO$_2$ is induced by CH$_3$OH-rich ice is now observationally supported by these three objects. From the measurements of these absorption bands in near- and mid-IR spectra, we computed a mean ratio of the column densities, $N {\rm (CH_3OH)}/N {\rm (CO_2)}\approx3$.

Previously, \citet{pontoppidan:08} found that CO$_2$ in CH$_3$OH ice is highly dilute in low- and high-mass YSOs in the Galactic disk, having a range of a mixing ratio from $N {\rm (CH_3OH)}/N {\rm (CO_2)}\sim3$ to $\sim20$. The diversity of the CO$_2$ concentration in CH$_3$OH ice could reflect differences in the evolutionary stages and/or envelope masses of these objects \citep[e.g.,][]{boogert:15}. Nonetheless, if the relatively low mixing ratio of CH$_3$OH:CO$_2$ in this study is taken at face value, it indicates that the CMZ clouds generally have a higher concentration of CO$_2$ in CH$_3$OH-rich ice than those found in the Galactic disk. Because CO$_2$ molecules can form by oxidation of CO, our result implies that some of CO in the grain surface have been converted into CO$_2$ rather than forming CH$_3$OH by a higher rate of oxydation. This may well be the case in the CMZ, owing to higher metallicity and oxygen abundance in the central region of the Milky Way \citep[e.g.,][]{carr:00,ramirez:00,schultheis:20}.

As a consequence, the characteristic $15\ \micron$ CO$_2$ absorption band of red CMZ objects in Paper~I can be naturally explained by a large concentration of CO$_2$ in CH$_3$OH-rich ice. Their band profiles are broad owing to a strong shoulder absorption component at $15.4\ \micron$, which is induced by Lewis acid-base interaction of CO$_2$ with CH$_3$OH \citep{dartois:99b}. Therefore, a large number of CO$_2$ molecules inside a CH$_3$OH ice matrix (as inferred from the above mixing ratio) would make the $15.4\ \micron$ absorption stronger than the other CO$_2$ components.

Nonetheless, the number of objects with direct abundance measurements of CH$_3$OH is small, because the absorption band at $3.535\ \micron$ is intrinsically weak. Another caveat in this study is that foreground YSO envelopes and/or dense cloud cores are traced by random lines of sight toward background stars, while the extended envelope of a massive YSO shows stratification of ice compositions as a function of the projected distance from the center \citep{vandertak:00,pontoppidan:04,cuppen:09}. As a consequence, the mixing ratio of CH$_3$OH and CO$_2$ ices can depend on which part of an isolated cloud is probed by background stars.

\subsection{Suppressed Star Formation in the CMZ}

One of the main findings in this paper is that there are only a handful of sight lines with significant water- and methanol-ice column densities. The sample observed in 2017 is particularly informative, because the number of sources in this sample is large, and they were selected based on a relatively simple selection function. From this, we found that there are only two lines of sight with strong water-ice absorption, if the selection cut on $N {\rm (H_2O)}$ is made based on objects in Paper~I having large water and methanol column densities and gas-phase absorption from warm and dense gas (\S~\ref{sec:ice}). This yields $2/68 = 0.03$ for a detection rate of intervening dense molecular clouds. The mid-IR color ($[8.0]\, -\, [24]$) excess of these two objects is not as high as those of the Paper~I sample, implying that foreground objects are in the later evolution stage of the Class~I YSO with a shallower spectral energy distribution \citep{lada:87}. In this regard, a small number of ice-rich sight lines found in this study resemble the paucity of dense cloud cores from millimeter and sub-millimeter observations \citep{kauffmann:17}.

Because ice-rich sources detected in this study represent very early stages of stellar evolution, their number can be used to constrain the current star-formation rate in the CMZ. The total expected number of foreground YSOs and/or dense clumps can be estimated in the following way based on the IRAC point-source catalog of the CMZ \citep{ramirez:08}. In this work, we selected the 2017 sample by imposing a combination of near- and mid-IR color cuts for the brightest objects in $[3.6]$ (Figure~\ref{fig:cmd}), but $[3.6]\, -\, [8.0] > 0.46$ was essentially the main criterion to select red point-like sources. Since there are approximately $37,000$ stars with the same $[3.6]\, -\, [8.0]$ cut in \citet{ramirez:08}, we expect a total $N=37,000\times(0.03\pm0.02)\sim400$--$2000$ ice-rich lines of sight, which pass through YSO envelopes and/or dense cores. The masses of these foreground objects are independent of the luminosity of the background sources, because they trace random sight lines. A cross-section of this random encounter increases with a size (therefore a mass) of a foreground YSO envelope, but the number of such massive YSOs is small as well. If we assume a standard mass function \citep[e.g.,][]{kroupa:01} and an order of magnitude larger radii of massive YSOs \citep{vandertak:00,pontoppidan:04}, these two factors are likely canceled out. Therefore, we can assume that a total cross-section by massive YSOs in the CMZ fields is comparable to those of lower-mass YSOs.

If we further assume an {\it average} mass of central objects to be $\langle M \rangle \la 1\ M_\odot$ and $\Delta t \ga 10^5$~yr for the lifetime of Class~0/I YSOs \citep[e.g.,][]{beuther:07,evans:09}, the star-formation rate becomes $N \langle M \rangle / \Delta t \la 0.004$--$0.02\ M_\odot\ {\rm yr}^{-1}$. This value is lower than previous estimates, $0.05$--$0.14\ M_\odot\ {\rm yr}^{-1}$ \citep{yusefzadeh:09,an:11,immer:12,longmore:13,nandakumar:18}. Therefore, despite large uncertainties, our result underscores the observed lower star-formation activity than predicted from nearby star-forming regions in the Galactic disk \citep{longmore:13,kruijssen:14,kauffmann:17}. 

Clearly, taking a full census of ice-rich sight lines in the CMZ is a key for understanding the star-formation activity and chemical evolution of molecular clouds. In this regard, future large IR spectroscopic surveys, such as the Spectro-Photometer for the History of the universe, Epoch of Reionization, and ices Explorer \citep[SPHEREx;][]{crill:20}, can provide an unbiased view on ice-rich clouds in this region. Large ground- and space-based telescopes with IR spectrographs will also be a valuable resource for following up these sight lines.

\acknowledgments

We are indebted to supporting scientists and telescope operators at the IRTF. We thank Thomas Geballe and Jeong-eun Lee for useful comments. D.J.\ and D.A.\ acknowledge support provided by the National Research Foundation (NRF) of Korea (No.\ 2018R1D1A1A02085433, No.\ 2021R1A2C1004117).

This publication makes use of data products from the Two Micron All Sky Survey, which is a joint project of the University of Massachusetts and the Infrared Processing and Analysis Center/California Institute of Technology, funded by the National Aeronautics and Space Administration and the National Science Foundation. This work is based in part on archival data obtained with the Spitzer Space Telescope, which was operated by the Jet Propulsion Laboratory, California Institute of Technology under a contract with NASA. Support for this work was provided by an award issued by JPL/Caltech.\\

\end{document}